\documentclass[sigconf]{acmart}  %%anonymous, authordraft
\usepackage{subcaption, algorithm, algorithmic, float, multirow, makecell}
\AtBeginDocument{%
  \providecommand\BibTeX{{%
    \normalfont B\kern-0.5em{\scshape i\kern-0.25em b}\kern-0.8em\TeX}}}

\copyrightyear{2023}
\acmYear{2023}
\setcopyright{rightsretained}
\acmConference[e-Energy '23]{The 14th ACM International Conference on Future Energy Systems}{June 20--23, 2023}{Orlando, FL, USA}
\acmBooktitle{The 14th ACM International Conference on Future Energy Systems (e-Energy '23), June 20--23, 2023, Orlando, FL, USA}
\acmDOI{10.1145/3575813.3595197}
\acmISBN{979-8-4007-0032-3/23/06}
\begin{document}
%\tableofcontents

%\title{Adapting Large Datacenter Loads to Reduce Grid Carbon Emissions}
\title{Adapting Datacenter Capacity for Greener Datacenters and Grid}
% Adapting Datacenter Capacity to Reduce Datacenter and Grid Carbon Emissions
% Adapting Datacenter Capacity for Greener Datacenter and Grid
%\title{Power Adaptation for Greener Datacenters under Grid-coupled Settings}
% How to Adapt Datacenter Loads to Reduce Grid Carbon Emissions Effectively
% we are actually seeking grid carbon reduction
% think about it from problem description
%\title{Datacenter vs. Grid: Exploring Load Adaptation Coordination to Reduce Operational Carbon Emissions}

%%Datacenter Load Adaptation in Harmony with the Power Grid}
% Synergistic Datacenter Load Adaptation: Sharing Information with the Grid is Key
% center phrase: datacenter load adaptation
% attribute: sustainability, grid-friendly

\begin{abstract}
Cloud providers are adapting datacenter (DC) capacity 
to reduce carbon emissions.  With hyperscale datacenters exceeding 100 MW individually, and in some grids exceeding 15\% of power load, DC adaptation is large enough to harm power grid dynamics,  increasing carbon emissions, power prices, or reduce grid reliability.  

To avoid harm, we explore coordination of DC capacity change varying scope in space and time. In space, coordination scope spans a single datacenter, a group of datacenters, and datacenters with the grid.  In time, scope ranges from online to day-ahead. We also consider what DC and grid information is used  (e.g. real-time and day-ahead average carbon, power price, and compute backlog). For example, in our proposed PlanShare scheme, each datacenter uses day-ahead information to create a capacity plan and shares it, allowing global grid optimization (over all loads, over entire day).

We evaluate DC carbon emissions reduction.
Results show that local coordination scope fails to reduce carbon emissions significantly (3.2\%--5.4\% reduction). Expanding coordination scope to a set of datacenters improves slightly (4.9\%--7.3\%). PlanShare, with grid-wide coordination and full-day capacity planning, performs the best. PlanShare reduces DC emissions by 11.6\%--12.6\%,   1.56x--1.26x better than the best local, online approach's results. PlanShare also achieves lower cost.  We expect these advantages to increase as renewable generation in power grids increases.  Further, a known full-day DC capacity plan provides a stable target for DC resource management. 

\end{abstract}

\author{Liuzixuan Lin}
\affiliation{
  \institution{University of Chicago}
  \streetaddress{5730 S Ellis Ave}
  \city{Chicago}
  \state{IL}
  \country{USA}
  \postcode{60637}
}
\email{lzixuan@uchicago.edu}

\author{Andrew A. Chien}
\affiliation{
  \institution{University of Chicago \& Argonne National Lab}
  \streetaddress{5730 S Ellis Ave}
  \city{Chicago}
  \state{IL}
  \country{USA}
  \postcode{60637}
}
\email{aachien@uchicago.edu}







%\tableofcontents
%\settopmatter{printfolios=true}
\settopmatter{printacmref=false}
\maketitle

%%\newpage
\section{Introduction}

%%motivation

With the commercial success of internet-scale applications and cloud computing, cloud infrastructure has grown rapidly. %%surpassing 100's of sites and continuing to grow at a prolific rate. 
A recent article documented the addition of over 50 datacenters a year by a single cloud provider \cite{Microsoft-50DC}.  %Rapid growth of cloud provider revenue and documented power purchases confirm 
By revenue, Amazon, Microsoft, and Google's cloud growth rates have exceeded 30\% annually for the past 5 years \cite{msftSustainability2021,googleRenewPurchase,Microsoft-30pct}.  The accelerated digitalization since COVID-19 \cite{Covid-Transform20}, and an accelerating adoption of machine learning (aka artificial intelligence) are both driving an acceleration of datacenter growth \cite{GreenAI-CACM20, genAICost}.  In 2021, the power consumption of these three cloud providers exceeded 62 TWh \cite{msftSustainability2021, amazonEnv, googleEnv}, equivalent to the power consumed by 6.2 million American homes.  In 2022 they purchased 14 GW of renewable generation capacity, but not enough to offset their power use \cite{ppa2022}.  Some estimates project that datacenter power consumption will grow %Accounting for 1\% of worldwide power consumption today, datacenters could 
to 10\% of global electricity use by 2030 \cite{Nature18,Masanet20, lin2021evaluating}.
%\aac{write sentences about total consumption of Amazon 30.9 TWh (2021), then Msft 13 TWh (2021) Google 18.3 TWh (2021) in TWh}  
%\aac{comment about projected fraction of world's electricity \cite{Nature18}}
Today's largest hyperscaler sites are multiple buildings with total power of 200 MW to 1 GW \cite{Greenpeace-Nova19,Microsoft-50DC,MicrosoftCloudDatacenters,AmazonCloudDatacenters,GoogleCloudDatacenters}.  

In many power grids, datacenters are already major load contributors.  In Virginia, datacenters account for 12\% of power consumption (2022), and will reach 18\% in 2027 and 22\% in 2032 \cite{dominion20IRP, dominion21IRP}.  In Ireland, datacenters  account for 14\% of national electricity use (2022) \cite{irelandDC14percent} and may be 30\% by 2029 \cite{irelandDC}.  With continued cloud and artificial intelligence growth, datacenters are expected to exceed 10\% or even 20\% of load in many power grids \cite{Nature18,Masanet20,dcMarketReport}.

Rapid computing growth raises concerns about carbon emissions \cite{Greenpeace-Nova19}.  Cloud providers often purchase renewable power (long-term contracts) or renewable offsets (renewable energy credits---RECs) to ``offset'' their power use \cite{googleRenewPurchase,microsoftEnv}.   However, these contracts are accounting arrangements, not actual power transfers.  Despite such arrangements and even full offsetting, cloud datacenters consume large quantities of fossil-fuel generated power \cite{bashir2021enabling, Greenpeace-Nova19}.  Worse, the growth of datacenters can threaten grid stability, 
%, which is reflected in power grid operators' recent actions. For example, worried that datacenter load growth may retard grid decarbonization and disturb grid operation, 
blocking DC projects in Ireland and Northern Virginia's grids. \cite{IrelandHaltDC, NoVAHaltDC}.

\begin{figure}[h]
    \centering
    \includegraphics[width=\columnwidth]{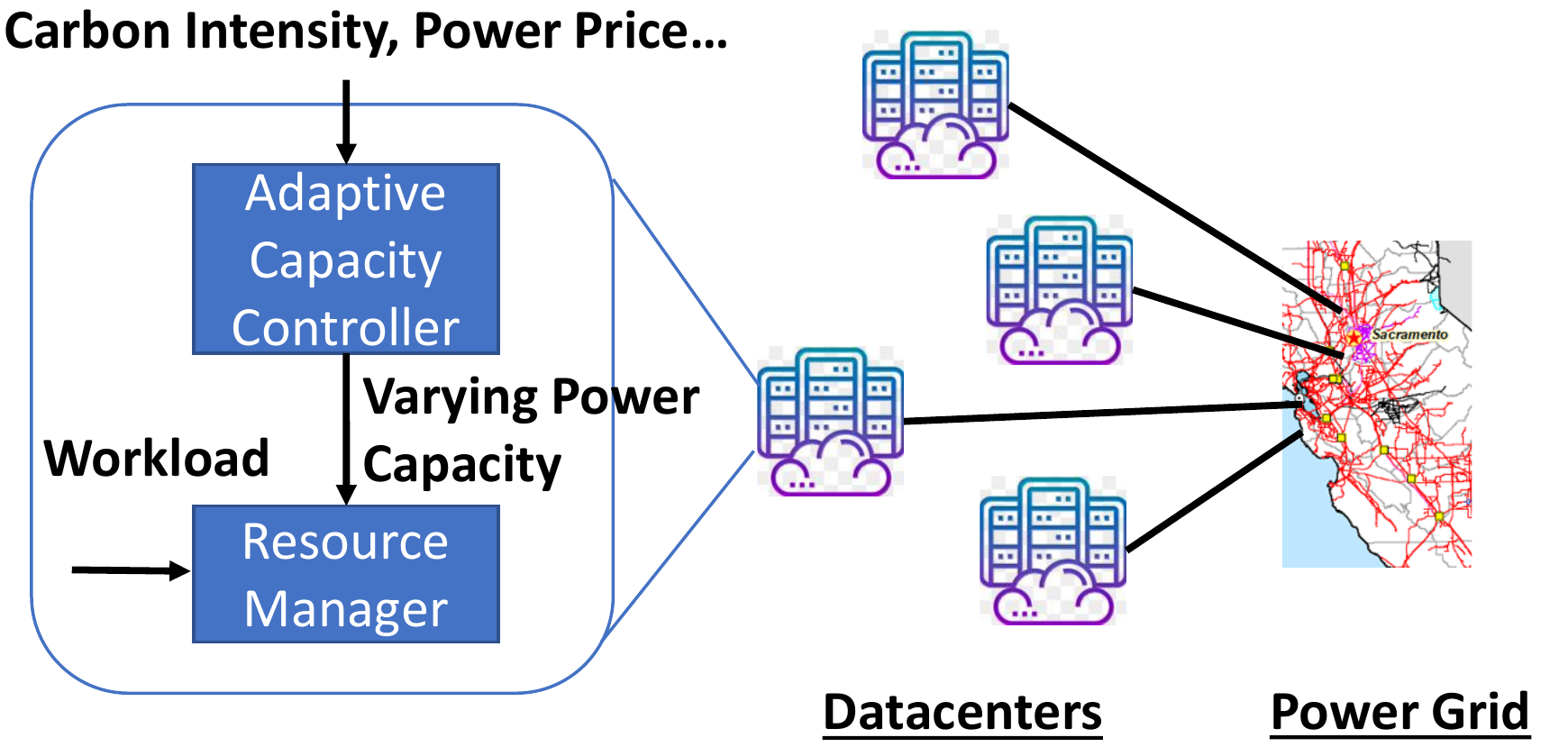}
    %\vspace*{-0.1in}
    \caption{Datacenters can adapt power capacity to align compute load with periods of lower carbon emissions.}
    \label{fig:intro}
    \Description{Within datacenters, the adaptive capacity controller controls power capacity according to power grid dynamics, and the resource manager makes decisions on workload using the power capacity information.}
\end{figure}

Adapting capacity (temporal shifting) to reduce DC carbon emissions has been explored for over a decade 
%datacenter carbon emissions (and power cost).  
%Changing location (geographic shifting) 
%(and also power cost from cloud provider perspective) of datacenters is capacity adaptation.  
%This reflects the goal of aligning power consumption with renewable generation.  
%And for Over a decade, computer systems researchers have proposed many creative ideas for dynamic load shaping across Internet datacenters with the goal of reducing carbon emissions 
\cite{goiri2013parasol,liu2012renewable,dou2017carbon, deng2014harnessing}, and to reduce power cost \cite{lin2012dynamic, liu2013data, luo2013temporal, Rosenthal19, Lin19, shi2016leveraging, urgaonkar2011optimal}.  For example, cloud providers' ``24$\times$7'' goals to match hourly power use and renewable generation typically involve temporal shifting \cite{GoogleWhite18,googlePolicyRoadmap,microsoftEnv,radovanovic2021carbon, acun2022holistic}. These approaches all seek to align compute load with low-carbon or low-price power, using online control techniques (see Figure \ref{fig:intro}).  Such efforts along with those we propose in this paper are increasingly crucial as power grids around the world driven by aggressive public policy are rapidly decarbonizing by adding new renewable generation \cite{NY70by30,EU-RPS45,Paris2015, California-rps-90}. %California-50RPS,
%for each day.has proposed a sophisticated adaptive load control to align datacenter power consumption with renewable generation  

%Some more aggressive efforts seek to increase the renewable absorption of the power grid \cite{KYZC2016,Lancium}.  

% summary of some prior related work
% - lots of work on edge control
% - lots of work on adaptive scheduling

%Recently, several cloud operators have explored research and pilot production efforts that explore application and resource manager level load shifting \cite{Manne20,radovanovic2021carbon}, but these efforts involve only small quantities of power, and therefore have correspondingly smaller carbon emission impacts.

%Recent work in e-Energy'21 shows that the combination of large datacenter load and load adaptation requires 

%%In short, growing cloud datacenter power demand is a top-shelf concern for power grids around the world, challenging grid operation and decarbonization. From the cloud provider perspective, 

DC shifting efforts typically 
%shape power consumption or shift load to reduce carbon emissions mainly 
focus on internal challenges of DC resource management.  
%and compute load prediction.  
A recent system forces shifting by creating a daily compute capacity plan to enable compute resource management  \cite{radovanovic2021carbon}.  However, most shifting efforts ignore the negative impacts of varying datacenter capacity on grid dynamics (e.g. triggering unnecessary generator starts and load shedding).  With large datacenters, such 
% the coupling of power capacity adaptation and grid 
impacts are critical.
%: 10-20\% of grid load is large enough to alter power grid dynamics.  
Load shifting schemes that are productive with small datacenters can be ineffective 
%fail to deliver benefits (carbon emissions, power cost) 
and even damage grid performance if used with large datacenters.  In Section \ref{sec:problem} (Problem), we show an example where such approaches create an 8\% increase in datacenter carbon emissions due to overshifting.   One consequence of this new insight is that there is no known solution to coordinate large-scale datacenter capacity adaptation and power grids.  To explore this new problem, researchers have begun to study capacity adaptation coupled with power grid models \cite{lin2021evaluating, menati2023modeling,lindberg2021guide}.  

To avoid the harm and effectively reduce DC carbon emissions with capacity adaptation, we explore coordination of DC capacity adaptation that varies in coordination scope---space and time. In space, we consider spans of a single datacenter, a group of datacenters, and datacenters with the power grid.  In time, we consider online and day-ahead.  We also vary datacenter and grid information used (e.g. real-time and day-ahead  metrics, compute backlog). For example, in PlanShare, each datacenter uses day-ahead grid information to create a full-day capacity plan, and then shares it with the grid, so the grid can optimize generation and transmission with a known, but varying datacenter capacity schedule.  
%use enables the power grid to optimize generation and transmission schedule across locations and time periods.
%First, we consider a widely-studied class of local datacenter, online intelligent control approaches. Second, to reduce emergent overshifting, we add coordinator(s) across DCs that moderates extreme collective behavior.  Finally, we consider a novel approach that adapts day-ahead based on forecasts, and shares plans with the grid, giving it warning to manage the changes.  %information sharing between DC and grid, 
%
%, and grid dynamics. Results from a global grid optimization, which maximizes grid benefits with DC load flexibility and also achieves low DC carbon emissions and power cost, are regarded as potential benefits to measure the effectiveness of these approaches. 
%
Our evaluation reports datacenter carbon emissions reduction relative to the fixed DC capacity scenario,
%(grid carbon emissions reduction attributed to capacity adaptation)
as well as cost impacts on both datacenters and other customers.

Specific contributions of the paper include:

%\vspace*{-0.2in}
\begin{itemize}
\item With local coordination scope, we use three grid metrics (average carbon intensity (ACI), grid price (Price), and locational marginal price (LMPrice)) for capacity adaptation. LMPrice is most effective, reducing datacenter carbon emissions 1--5\% vs. ACI and 0.7--1.5\% vs. Price.  %A further advantage is that LMPrice is practically available. 
Exploiting hourly future price information (+2.3\%) and step size (+2.3\%) further improves local adaptation based on LMPrice, achieving datacenter carbon reduction of 10\%.

\item Expanding the coordination scope to a group of datacenters with a coordinator that limits aggregate behaviors gives only a small improvement over local adaptation, reducing datacenter carbon emissions by 7.3\%.
%We also consider a coordinator to limit aggregate load change. This improves LMPrice-based adaptation, giving a carbon reduction of 7.3\%.  We next consider how future prices (+2.3\%) and step size (+2.3\%) can increase the effectiveness of local adaptation based on LMPrice, this is a greater benefit than with coordinator and reaches a total carbon emissions reduction of 10\%.  %  Multiple datacenters given correlated grid signals, can adapt together, producing large, rapid changes in load that cause grid harm.  

%%To maintain datacenter load adaptation control, and recognizing that unexpected load changes cause grid problems, we consider approaches where datacenters share their planned load adaptation 1--24 hours in advance.  This enables the grid to plan ahead and produces much better results\footnote{Note that Google's Carbon-aware platform builds a 24-hour plan, but does not share it with the grid \cite{radovanovic2021carbon}.}. 
%Finally, we consider a cooperative scheme, where datacenters share their full-day adaptation plan with the grid in advance.  
\item PlanShare, datacenter-grid coordination and full-day capacity planning, achieves  greatest benefits, decreasing datacenter carbon emissions by 11.6\%--12.6\%. This grid-wide coordination is 
% with realistic wind penetration levels
1.56x--1.26x better than the best local, online approach. The key costs (grid dispatch cost, customer power cost) are also reduced. Further, for datacenters, the 24-hour capacity plan provides a stable target enabling more efficient compute resource management.  PlanShare 
%is a widely usable solution that 
satisfies datacenter and power grid objectives.

%for the overall grid and other customers.

%The shared schedule enables the grid to perform better global optimization.  A sensitivity shows the benefits decrease the length is decreased to 12, 6, etc. hours.

%warning cooperation increases the benefits

%by 10.2\% to 12.6\%, corresponding to a reduction of 1.45 million metric tons CO$_2$ (equivalent to the annual power-carbon footprint of 380,000 US homes). 

\end{itemize}

%% rest of paper organization
The remainder of the paper includes Background (Section \ref{sec:background}) and in Section \ref{sec:problem} we describe the challenge of adapting large-datacenter capacity.  In Section \ref{sec:approach} we discuss datacenter capacity adaptation approaches.  Section \ref{sec:methods} introduces the methodology of grid-coupled simulation. In Section \ref{sec:evaluation} we evaluate carbon reduction of the different capacity adaptation approaches, considering impacts on datacenter and non-datacenter grid customers. Finally, in Section \ref{sec:related} and \ref{sec:summary}, we discuss related work, summarize results, and discuss future research directions.

\section{Background}
\label{sec:background}

\subsection{Growth of Cloud Datacenters in Power Grids}
Hyperscale cloud providers (e.g. Amazon, Microsoft, Google, Alibaba) are building larger and more datacenters to meet the needs of digitalization, which accelerated since the COVID-19 pandemic \cite{Covid-Transform20, Covid-Transform20a, dominion21IRP}. The rapid growth of hyperscale cloud industry is reflected in its power consumption: 1\% of worldwide power consumption in 2018, projections suggest it could exceed 10\% by 2030 \cite{Nature18,Masanet20, lin2021evaluating}.  Similar rapid growth rates are also supported by corporate renewable power purchase data \cite{googleRenewPurchase,msftSustainability2021}. 
%Some regions like Northern Virginia or Ireland already see high fraction of datacenter load (higher than 10\% today and may reach 20\%--30\% in 10 years ) \cite{dominion20IRP, irelandDC}.

%Another example is Ireland, where datacenters use 14\% of national electricity use today \cite{irelandDC14percent} and could use 30\% by 2029 \cite{irelandDC}. Worried that datacenter load growth may hold back grid decarbonization, Ireland's state grid operator is canceling datacenter projects \cite{IrelandHaltDC}.

As a result, datacenters have become the key driver of new electric power demand, direct cause of new emissions, as well as construction of power plants, transmission, and energy storage infrastructure \cite{dominion20IRP, dominionMoreGreen, NoVAHaltDC}. In regions where datacenters account for large fraction of power consumption (e.g. Virginia---12\% in 2022 and 22\% in 10 years, Ireland---14\% in 2022 and 30\% by 2029), the challenges in grid operation and decarbonization have become evident. For example, Ireland's state grid operator is canceling datacenter projects \cite{IrelandHaltDC} unless the datacenters bring their own renewable generation or adapt their power demand. Similar story is happening in more and more parts of the world as the scale and reach of datacenters continue to grow \cite{NoVAHaltDC, Nature18,Masanet20,Greenpeace-Nova19,dcMarketReport}.  
%In short, growing cloud datacenter power demand is a top-shelf concern for power grids around the world, challenging grid operation and decarbonization.

%growth of size of datacenters

%growth of total cloud datacenter (google and microsoft datacenter purchases \cite{googleRenewPurchase})

%amazon is the largest, and not yet doing this (we have a lower bound)

%examples of where the datacenters are a large part of the power grid load has >5\%, 10\% by 2030 in nova\cite{Greenpeace-Nova19}

%they will be increasingly so

%some cloud networks are beginning to do load shifting (google carbon aware), but the fraction and load are small and thus give small carbon emissions benefit
From the cloud provider perspective, such exponential growth in datacenter power consumption first translates into growing power cost, as power price \cite{eiaPowerPrice} don't decrease at the same speed. Also, it raises concern about carbon emissions growth. Obscured by carbon offset or renewable purchase, additional datacenters can still result in more fossil fuel consumption if the load is not aligned with renewable generation \cite{Dominion18IRP, irelandDC, bashir2021enabling}. To improve sustainability, Google and Microsoft seek to hourly match their datacenter power capacity with renewable generation (24$\times$7, 100/100/0 \cite{GoogleWhite18, msftSustainability2021}), with attempts such as capacity adaptation \cite{radovanovic2021carbon}. While the current scale of load change is only a few percent of capacity \cite{radovanovic2021carbon}, it is expected to increase in future (discussed in Section \ref{subsec:flexibility}).  % rapidly due to the sustainability commitments and increasing incentive from grid decarbonization (see below). 
Furthermore, such adaptation behaviors can disturb the grid even at a small percentage of capacity. There are already reports of power variations from supercomputers (all on or all off) affecting grid stability \cite{Supercomputer-Unstable-Grid19}. For gigawatt cloud datacenters, a 10\% load change produces a 100 MW swing, similar to the dynamic range we study; a 40\% load change, which is not unusual for some diurnal peak to trough, would be 400 MW!

%even at a small percentage the swings are growing, and there are already reports from supercomputer loads (all on all off) that can damage grid stability.  for cloud datacenters of a gigawatt, a 10\% load change would be a 100MW swing, approximately the dynamic range we study.  a 40\% load change, not unusual for some diurnal peak to trough, would be 400MW!

%as the incentive increases (see below), the percentage shifting will increase
% china data: IEA 2019, europe data: eurostat

\subsection{Dynamics of a Renewable-based Power Grid}

%Climate agreements seek to limit global warming to below 1.5$^{\circ}$C above pre-industrial to avoid catastrophic climate-related risk. This means the world's carbon emissions need to be halved by 2030 and net-zero by 2050 \cite{masson2018global}. Power grid decarbonization is necessary for achieving these goals. 
Aligned with the carbon reduction goals (e.g. halved by 2030 and net-zero by 2050 \cite{masson2018global}) that seek to limit global warming, recent years have seen the rise of renewable sources in energy mix of many power grids across the world, such as 34\% in California (2021), 27\% in China (2020), and 22\% in Europe (2020). There are more ambitious goals for this decade. For example, California aims at 60\% renewable fraction by 2030 \cite{rps-california}, and Germany plans to phase out coal power plants (30\% of electricity supply in 2021) by 2030 \cite{germanyPhaseOutCoal}.

Integrating intermittent renewable sources (mainly wind and solar) is challenging for the power grid. For example, renewable generation can be wasted due to temporal mismatch with energy demand and transmission limits, producing both negative-priced power (aka stranded power) \cite{caisoManagingOversupply,CAISO-stranded18,ERCOT-stranded20} and even``curtailment'' \cite{ERCOT-stranded20,CAISO-stranded18,AIMS18,Bird2013,GWEC-Annual16,ChinaWilson15} where the renewable generation is wasted (not used by the power grid). A complementary problem is generation shortage, such as under extreme weather (heat, storms) where wind or solar generation can be dramatically lower. One potential solution to this challenge is to increase supply or demand flexibility, and corresponding methods include energy storage and adaptive loads \cite{caisoManagingOversupply,LFLTF}.

\begin{figure}[h]
    \centering
    \includegraphics[width=\columnwidth]{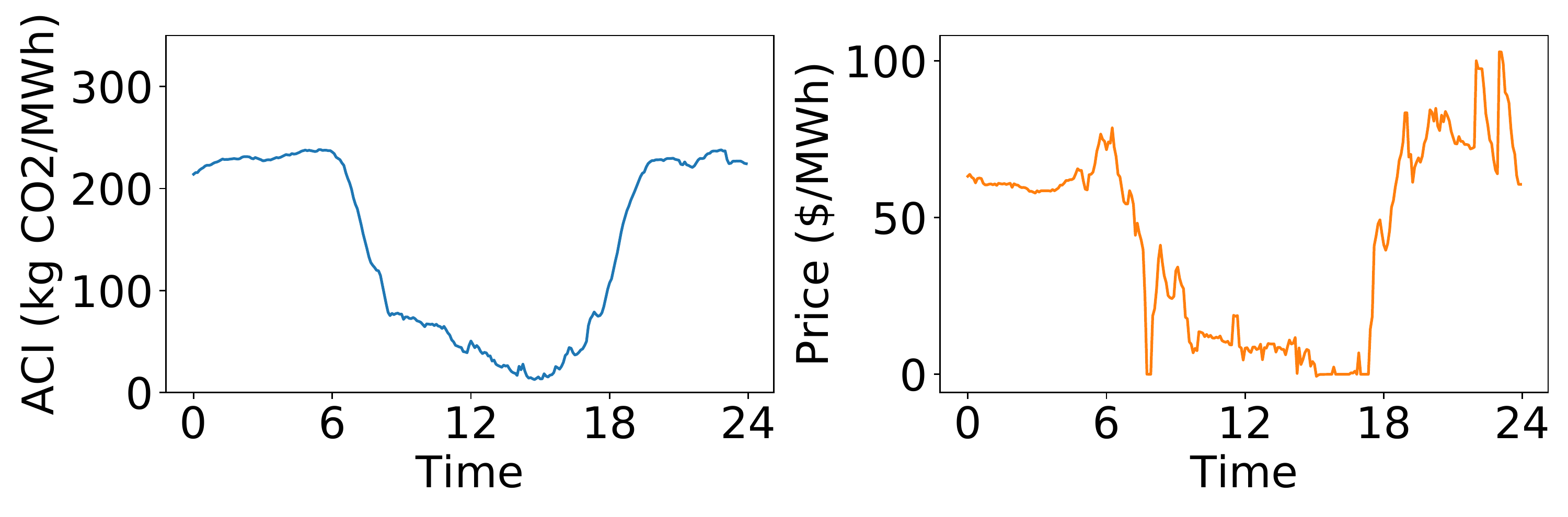}
    \caption{CAISO's Daily Average Carbon Intensity and Price Variation, 2022/05/02. Left: Average Carbon Intensity (kg CO2/MWh), Right: Grid Price (\$/MWh). Source: CAISO.}
    \label{fig:caiso_aci_price}
    \Description{Daily average carbon intensity and power price in CAISO, California's power grid are correlated. They are both in U-shape and the valley is shaped by solar generation.}
\end{figure}

The growth and intermittency of renewable generation produce time-varying grid metrics, such as carbon intensity (carbon emissions per MWh energy consumption) and power price. For example, in California's power grid, the average carbon intensity keeps low during the daytime when solar generation dominates, but it climbs up to about 200 kg CO$_2$/MWh as natural gas generators are up (Figure \ref{fig:caiso_aci_price}, left). In addition, as wind and solar generation is usually bid low due to zero fuel cost, the power price correlates with carbon intensity. The price can drop to near zero when there is excess renewable generation (Figure \ref{fig:caiso_aci_price}, right)! Adaptive loads such as datacenters, electric vehicles, and smart appliances may exploit such variation to reduce their carbon emissions or power cost.

\section{Problem}
\label{sec:problem}
%%different control objectives, cost of grid-controlled optimization $=>$ intelligent local control scheme
%5 \aac{edit to a clear problem focus \\ how to maximize datacenter RPS increase? \\ by adapting load profile}

Cloud datacenters seek to reduce carbon emissions by aligning power use with plentiful renewable generation (low-carbon power).  However, the carbon-intensity of grid power is determined by  complex interaction of load, transmission, available generators and their ramping rate-limits. Grid dynamics can be opaque because power grids are a critical infrastructure and power markets have fierce economic competition.  Together, these factors limit the real-time telemetry available to guide capacity adaptation.
%Besides, with current scale, datacenters' adaptation behaviors can alter the grid metrics they adapt to. 
Therefore,  it is challenging for large-scale cloud datacenters to 
dynamically choose capacities that reliably reduce carbon emissions.  Often, load shifts achieve only a fraction of anticipated benefits, or worse, they can harm themselves or other customers by increasing prices or carbon emissions.  
%For example, % and power cost, creating problems for the grid and limiting the benefits of increased renewables. This limitation is illustrated in 
%Figure \ref{fig:decrease-renewable} shows that while local, intelligent datacenter load adaptation can increase {\it datacenter RPS}, reducing the carbon emissions for the power the datacenter actually consumes, it only captures as little as 30\% of grid potential benefit.  In some cases, it can even decrease datacenter RPS.
%even retard renewable absorption.

%\aac{Do we need to make the case that DC RPS is the RIGHT metric?}

\begin{figure}[h]
    \centering
    \includegraphics[width=\columnwidth]{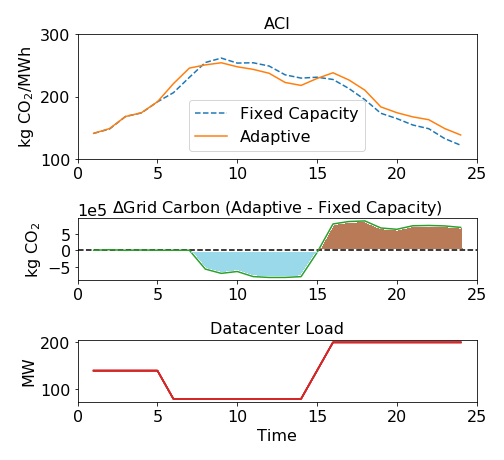}
    \caption{Overshifting: local, online adaptation using average carbon intensity (ACI) shown in the Top increases datacenter carbon emissions by 8\% (Middle) because they all make the same capacity adaptation (Bottom).}
    \label{fig:overshiftingExample}
    \Description{Datacenters' local, online adaptation using average carbon intensity (ACI) moves the grid average carbon intensity curve, increasing datacenter carbon emissions by 8\% because they all make the same capacity adaptation. We call this phenomenon "overshifting".}
\end{figure}

%\footnote{Independent DC Control is the Indep-ONL-Avg algorithm, and Potential Benefit is from GC in Section \ref{sec:methods}.}  
One example of this problem is ``overshifting''. 
%where a set of datacenters increase load together, triggering the dispatch of additional fossil-fuel generation and increasing carbon emissions.  
Simulating a Spring day with our grid model (Figure \ref{fig:overshiftingExample}),
with intelligent local control based on grid {\it average carbon intensity (ACI)} \cite{electricityMap, tomorrowCarbonIntensityBlog}, all datacenters decrease capacity at hour 5 (4 am) and increase capacity at hour 15 (2 pm), causing the entire grid's ACI to increase by a total equivalent to 8\% of datacenter daily carbon emissions (vs. fixed capacity)!  Why is overshifting a problem? The local capacity control of the datacenters reacts to a shared metric in unison, increasing or decreasing capacity at the same time (2nd row).
%The adaptation reference is what would have happened if the datacenters operated at fixed capacity.
%of grid constraints (e.g. generation, transmission, ramping), act independently but in unison, 
The large capacity changes, combined across datacenters, oversubscribe the opportunity.  In order to maintain grid balance, the market responds by dispatching additional generation (fossil-fuel!), and thereby increasing carbon emissions.

%as shown in Figure \ref{fig:overshiftingExample} (Summer Weekday, 15\% wind).  

%The root cause for the general shortfall and overshifting is conflicts between datacenter load adaptation and grid-wide optimal power flow (OPF) optimization.
%Under regular operation, power grids coordinate generation and transmission spatially and over the 24-hour day.  They use generation flexibility to smooth overall grid generation-load balance, avoiding dispatch of high-cost generators and penalties for over- or under-generation.  This smoothing benefits all loads in the grid, including datacenters, but is not possible if datacenters change load abruptly under independent control. 

Thus the challenge for datacenters is: \textbf{How to adapt capacity to reduce operational carbon emissions?}  That is, how to align power use with low-carbon opportunity without disturbing the grid. To achieve this, capacity adaptation must: 

\begin{enumerate}
    \item Identify opportunity: how to find the right times to increase and decrease capacity to reduce their carbon emissions?
    %from available grid metrics. For example, when renewable generation increases faster than load, or when load decreases faster than renewable generation.
    \item Avoid contention or overshifting: how to share the opportunity with other potential adaptive loads in the grid? %of reducing carbon emissions, datacenter power adaptation 
    (avoid oversubscribing the opportunity)
    \item Avoid harming others: how to ensure capacity adaptation does not harm others---load participants (consumers, other companies, other datacenters) by increasing prices or carbon emissions, or generators and grid resilience?
%    have competing interests, so there is no global optimum. However, harming other (non-DC) customers (e.g. increasing average power price) is obviously not a goal of datacenter power adaptation.%, which prior study has shown can happen.
\end{enumerate}

%Important challenges include avoiding the damages caused by overshifting or other sudden load changes, and how to maximize the generation mix benefits of the load flexibility.

%This underscores the fact that many power grids operate as open power markets with direct financial control and competition. Clearly, such effects should be minimized.  

%%challenges: edge control, grid response (ramp rates, schedule)

\section{Approach}
\label{sec:approach}

\begin{figure}[h]
    %\centering
    \includegraphics[width=\columnwidth]{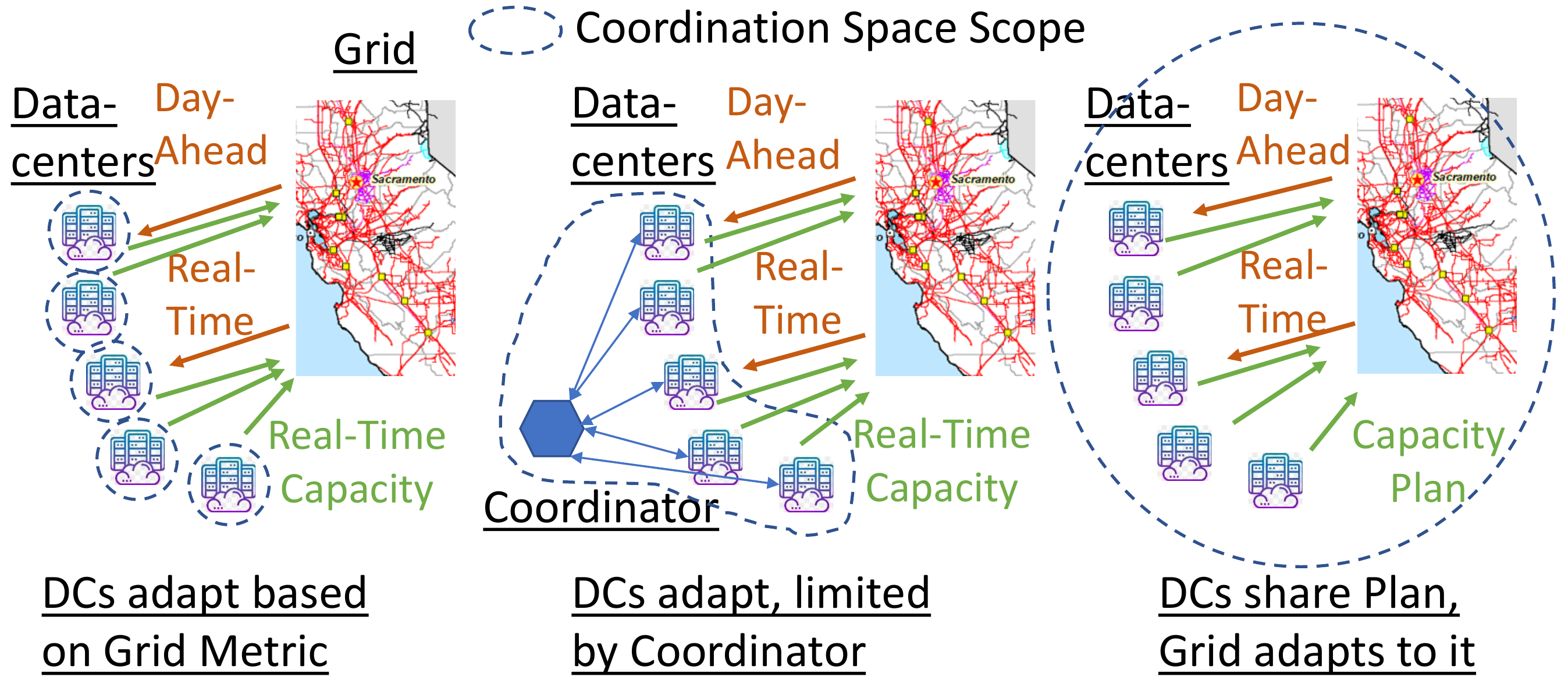}
    %%\textbf{need a picture to show these three approaches}
    \caption{Adaptation Approaches vary in Coordination Scope.}
    \label{fig:approaches}
    \Description{The three approaches in this figure vary in two dimensions of coordination scope: space and time. The space ranges from a single datacenter to a group and then to datacenters with the grid.  The time includes online (real-time) to day-ahead (24 hours). The datacenter and grid information used for coordination is also varied.}
\end{figure}

To address the challenges, we explore different datacenter capacity adaptation approaches and couple them to grid simulations to evaluate the \textbf{realized impacts}.
The approaches vary in two dimensions of coordination scope: space and time. The space ranges from a single datacenter to a group and then to datacenters with the grid.  The time includes online (real-time) to day-ahead (24 hours).  In Figure \ref{fig:approaches}, we illustrate three spatial scopes of coordination.   We also vary the datacenter and grid information used for coordination.

%\aac{local control,online}
Representing a classes of control techniques in previous work, \textbf{local} adaptation (Figure \ref{fig:approaches}, left) makes hourly decisions using real-time and future metrics from the grid.  Our studies of local adaptation  show the limitations of these approaches.  But first, we compare grid metrics (e.g. average carbon intensity, power price) that datacenters can use to drive local capacity adaptation.  

%to. In addition, we explore the improvements from future grid information and smoothed adapted datacenter capacity with step size.  % control of datacenters.

%\aac{above + coordination across multiple datacenters}

At times, local adaptation can increase carbon emissions (e.g. overshifting in Figure \ref{fig:overshiftingExample}).  We consider use of an external \textbf{coordinator}  that coordinates a group of datacenters to eliminate the harm (Figure \ref{fig:approaches}, middle).  In this scheme, 
each DC makes an hourly request to adapt capacity, and the external coordinator limits the group's total capacity change.  
%This smooths the load changes that the power grid must manage with to reduce negative grid impacts.  
%Some DCs will not get their requested level of change, and DC benefits may be reduced. 

%\aac{information sharing, day-ahead adaptation}

Finally, we consider a new approach that expands coordination scope to the entire grid.  In this approach each DC \textbf{plans ahead}---making a binding capacity plan 24 hours in advance based on forecast grid information, and \textbf{shares the capacity plan with the grid}.
The plan provides datacenter load certainty to the grid, enabling it to optimize generation and transmission scheduling under dynamic constraints that span single or many hours.
%over r  This approach empowers the grid's optimization to manage any planned changes.
%the base scheme, the DCs no longer make online decisions about power level, avoiding disruption of the 24-hour plan created with the grid's OPF algorithm.

%\aac{start with the metric of success! \\ then also talk about other considerations -- eg grid impact }

We introduce the algorithms for each approach, evaluating them in Section \ref{sec:evaluation}.  We report DC carbon reduction with realistic levels of wind generation and datacenter load fractions (3.5\%--14\%), framing what could happen today or in the next 5--10 years.  For clarity of exposition, we focus on a 30-DC scenario, representing about 10\% of grid load that approximates the current levels in Northern Virginia (12\%) or Ireland (14\%), and vary the wind penetration (ratio of average wind generation to grid energy demand) from 15\% to 60\% (2015 level--2050 target).  We also report how DC capacity adapatation creates grid impacts on power prices and how the impacts correlate with datacenter capacity change.

%The simulations are with realistic settings: datacenter sizes, power grid model with wind penetration level  development from today to the coming decades (15\%--60\%, 2015 levels--2050 target), and 10 to 40 datacenters, which match current and near-future (2027) loads in Northern Virginia and Ireland power grids. For clarity of exposition, we focus on a 30-DC scenario, representing  $\approx$10\% grid load, below current levels in NoVA's (12\%) and Ireland's (14\%) power grids.

%Each datacenter load has a maximum power of 200 MW and 70\% average utilization (140 MW) in the 24-hour period. These loads can be modulated within [60\%, 80\%] (narrow) or [40\%, 100\%] (wide) dynamic range. We studied from 10 to 40 datacenters, but focus on the 30-DC here for clarity ($\approx$10\% grid load).  %% when the tension between dynamic DC loads and grid becomes significant, especially with large dynamic range.

% describe them abstractly, with a few simple pictures?

% smart edge, smart edge with coordinator, individual datacenters and schedule sharing

% \begin{verbatim}
% - more sophisticated edge control (local)
% - step limits

% since the problem is the aggregate impact
% - coordinator, various schemes

% grid shares information about its prices
% and schedule - day-ahead, what if the datacenters
% did the same?
% - day ahead schedule from grid
% - optimization (static) of datacenter load
% - sharing of the schedule with the datacenter
% \end{verbatim}

\section{Datacenter and Grid-coupled Simulation Methodology}
\label{sec:methods}
This section describes the framework for evaluating the datacenter capacity adaptation approaches. Datacenters adapt capacity to grid metrics (Section \ref{subsec:gridMetrics}), respecting the capacity flexibility constraints (Section \ref{subsec:flexibility}). Resulting time-varying loads affect the dynamics (e.g. pricing, generation) in the power grid % The simulations are grid-coupled 
(Section \ref{subsec:gridModel}).

%so that we can examine the realized effects of load adaptation, with evaluation metrics in Section \ref{subsec:metrics}.

%%while we mainly focus on the overshifting problem.

%\subsection{DC-grid Coupling Models}
\subsection{Grid Metrics for DC Capacity Adaptation}
\label{subsec:gridMetrics}
%Because of 
Power grids have complex dynamics, so % and data availability, choice of 
the ``best'' grid metric\footnote{Sometimes, these are called grid ``signals'' for load adaptation.} for reducing carbon emissions is an open research question \cite{tomorrowCarbonIntensityBlog, lindberg2021guide}.  A good metric should enable carbon reduction and be available in most or all power grids.  We consider several candidates:

\begin{itemize}
    \item \textbf{Average carbon intensity or ACI (kg CO$_2$/MWh)} is the carbon emissions per MWh energy consumption in the grid. Derived from fuel mix, ACI is usually only available from unvalidated 3rd parties (e.g. Electricity Maps \cite{electricityMap}) in a fraction of the world's grids. % and for less than half of the world.
    \item \textbf{Grid price or Price (\$/MWh)} is power price in a grid or region (e.g. ``hub price''). Renewable generator often bids low, causing power price to be correlated with carbon intensity (Figure \ref{fig:caiso_aci_price}). Price information is widely available in day-ahead and real-time markets. % to market participants and in different processes of markets (e.g. day-ahead, real-time).
    \item \textbf{Locational marginal price or LMPrice (\$/MWh)} is the price at a specific node in the power grid.  %As a nodal signal, 
    LMPrice reflects local properties such %is more spatially fine-grained than the other two signals above, reflecting 
    as nearby renewables and grid transmission constraints. %and local accessibility of renewable generation. For example, when the grid sees low average ACI or GPrice, increasing load at a place with high LMPrice may fail to utilize that. 
    LMPrice is widely-available. 
    %too---GPrice is the average of LMPrice weighted by loads.
\end{itemize}

Recently, researchers and companies \cite{lindberg2020environmental,wattTimeAER} have proposed marginal carbon intensity as a metric. %---change of carbon emissions with respect to load. 
We do not consider it, as it is not widely available, and tied to proprietary market strategy.
%we don't consider it as it hasn't been broadly available.
%very limited, making it hard to assess. It's also sensitive to load change and not stable. With the goal of finding a widely usable solution, we don't consider it in this paper.

\subsection{Modeling Datacenter Capacity Flexibility}
\label{subsec:flexibility}
\begin{figure}[h]
    \centering
    \includegraphics[width=\columnwidth]{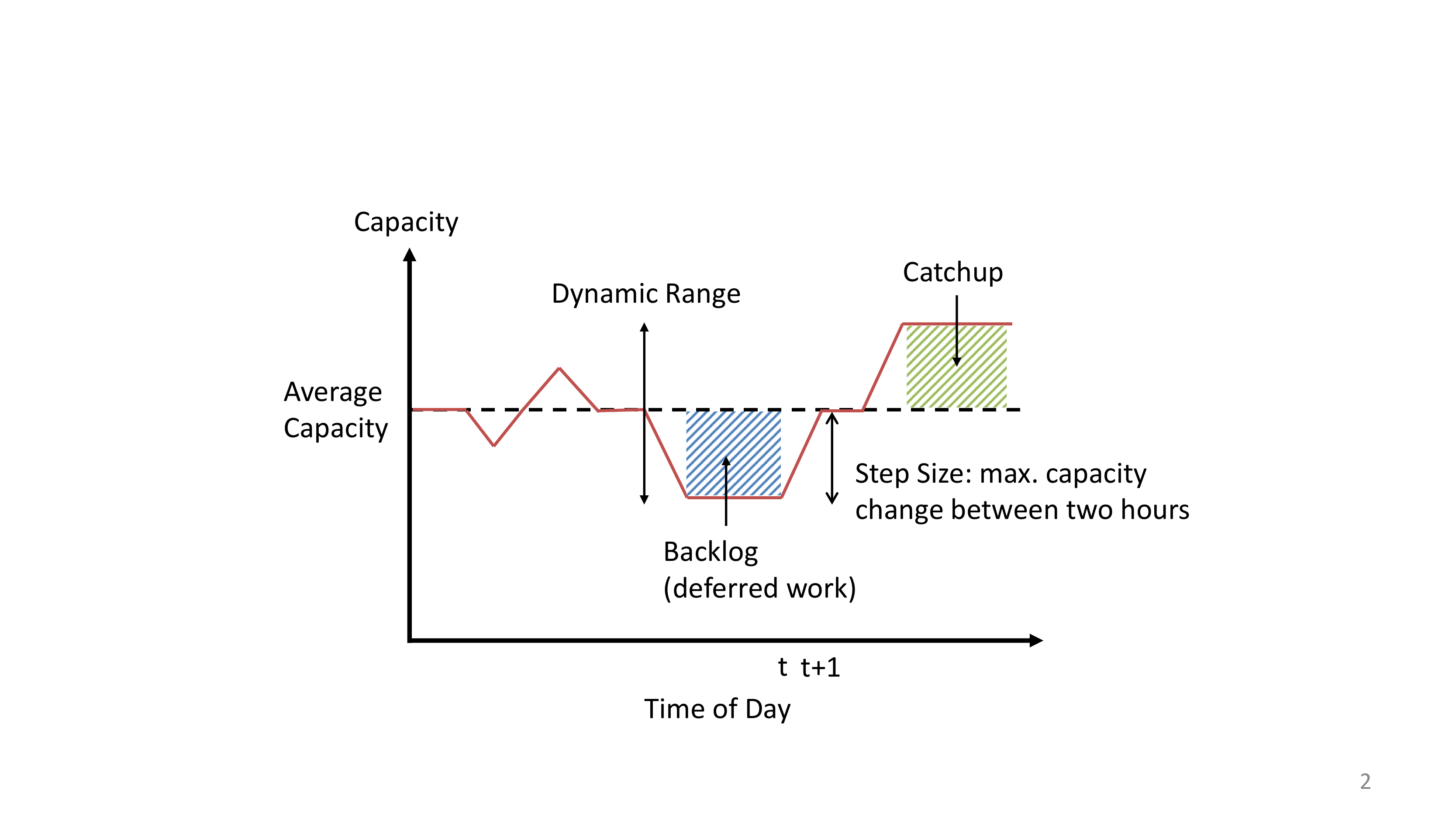}
    \caption{Datacenter Capacity Flexibility Model.}
    \label{fig:dc_flexibility}
    \Description{The capacity flexibility constraints are illustrated in a daily capacity example. The datacenter can adjust capacity within a dynamic range and defer workload, but must catch up within a 24-hour day. There is also step size limit that bounds hour-to-hour capacity change.}
\end{figure}

Datacenter capacity flexbility defines the variation structure of capacity. We assume that every datacenter can adjust capacity (\textbf{cap}) within a \textbf{dynamic range} and defer workload (\textbf{backlog}), but must catch up within a 24-hour day (Figure \ref{fig:dc_flexibility}), formally:
\begin{equation}
\label{cons:dynRange}
    cap_{min}\le cap_{i,t}\le cap_{max},\ \forall i,t
\end{equation}
%i.e. all datacenter loads $\{l_{i,t}\}$ are always within the dynamic range. At time $t$, datacenter $i$ updates its backlog as:
\begin{equation}
    backlog_{i,t}=backlog_{i,t-1}+(avgCap-cap_{i,t})
\end{equation}
\begin{equation}
    backlog_{i,24}=0,\ \forall i
\end{equation}
These flexibility constraints are common in datacenter carbon-aware capacity adaptation studies \cite{lin2021evaluating, acun2022holistic, radovanovic2021carbon}.

%Where The backlog is non-negative and must be zero at the end of the day to satisfy the average load constraint.

Datacenter capacity variation can harm a datacenter's \textit{computation performance} \cite{zhang2021scheduling} and even power markets. % a power grid's ramping capability.  
We consider a \textbf{step size} limit that bounds hour-to-hour datacenter capacity change:

%%to smooth individual datacenter's capacity, which limits 
\begin{equation}
\label{cons:stepSize}
    \lvert cap_{i,t}-cap_{i,t-1} \rvert \le stepSize,\ \forall i,t
\end{equation}

%\textbf{rewrite below as direct}
DC attributes used are shown in Table \ref{tab:sim_config}. The resource utilization and capacity assumptions are typical of hyperscale datacenters \cite{BorgTNG20,GoogleCloudDatacenters,MicrosoftCloudDatacenters,Greenpeace-Nova19}. We present results from the [0.4, 1.0] dynamic range which combines the largest potential benefits, but also overshifting challenges.  High fractions of deferrable workload reflect the industry's published workload papers.  Google's BorgTNG trace \cite{BorgTNG20} shows flexible jobs with 24-hour completion SLO (service level objective) make up about 40\% of the resource usage; delay-insensitive VMs (virtual machines) account for about 68\% of resource usage among Microsoft Azure VM workload \cite{ResourceCentral17}; in Meta, 60\% of batch jobs can be flexibly scheduled within a day \cite{acun2022holistic}. It's likely that the workload flexibility continues to grow due to emerging batch workloads like machine learning model training. In addition, the potential deployment of long-duration energy storage %long-lasting batteries 
is a complementary solution to datacenter capacity adaptation \cite{longLastingBattery}.

\begin{table}[h]
\caption{Configurations of Datacenter Attributes}
\label{tab:sim_config}
\begin{center}
\begin{tabular}{p{110pt}|p{110pt}}
  \hline
  Attribute & Configuration(s)\\
  \hline
  Maximum Capacity & 200 MW\\
  Average Utilization Level & 70\%\\
  Average Capacity ($avgCap$) & 140 MW\\
  Dynamic Range & [0.6, 0.8], [0.4, 1.0]\\
  Step Size & 10, 20, 40, 80, 120 MW/h\\
  \hline
\end{tabular}
\end{center}
\end{table}

%\textbf{make this comments someplace once?}  \cite{lin2021evaluating}

%Rarely, the coordinator can cause a DC's backlog constraints to be violated.  This happens if quota prevents DCs from increasing load to catch up.  The complementary problem is overconsumption, where the quota prevents DCs from decreasing power load.  Both of these situations are rare and the impact on DC capacity is small -- for backlog averages $<1\%$ and for overconsumption $0.1\%$.
%%and max of 1\% in our experiments.

%%datacenters can have workload backlog at the end of the day. Another case is when the quota prevents datacenters from decreasing load, overconsumption of power (negative backlog) can happen. In this case we set the backlog after that hour as 0 according to the assumption that workload can only be deferred.

%Such accurate future DC load information enables the power grid to perform global optimization---space and time---and benefit the datacenters in turn.

\subsection{Power Grid Model}
\label{subsec:gridModel}
We evaluate DC adaptation approaches coupled to a realistic grid model. Grid operation (day-ahead planning or real-time operation) is simulated by solving the direct-current optimal power flow (DC-OPF) problem in \cite{KYZC2016, lin2021evaluating} and Appendix \ref{appendix:dc-opf}, which minimizes the grid dispatch cost in one-day time horizon with hourly intervals, subject to typical grid constraints. The grid metrics for carbon optimization (ACI, Price, LMPrice) are derived from the OPF solutions. With lower generation costs and curtailment penalties that encourage use, renewable generators produce low prices when dispatched at the margin, capturing the the correlation between carbon metrics and power price in the real world (Figure \ref{fig:caiso_aci_price}).

The grid topology is a reduced California power system (CAISO) consisting of 225 buses, 375 transmission lines, 130 thermal generators (31.2 GW total capacity), 11 non-wind renewable power plants, 5 wind power plants, and 40 loads. Power can also be imported at 5 boundary buses. %The thermal power plants' ramp rates are scaled up by 4-fold to reflect current ramping capability of CAISO as in \cite{lin2021evaluating}. 
This model is originally from \cite{papavasiliou2013multiarea} and has been used to assess the impact of dynamic datacenter capacity management in \cite{KYZC2016, lin2021evaluating}. We select wind as the major renewable generation source as it presents more intra-day variation that leads to more diverse capacity adaptation behaviors. Besides, it can model the wind-dominant power grids such as ERCOT (Electric Reliability Council of Texas) and SPP (Southwest Power Pool). To get higher wind penetration, the wind generation are scaled up equally at current sites, assuming those sites can be expanded or equipped with higher-capacity wind turbines \cite{pryor202020}.

\begin{figure}[h]
    \begin{subfigure}{.27\columnwidth}
        \centering
        \includegraphics[scale=0.29]{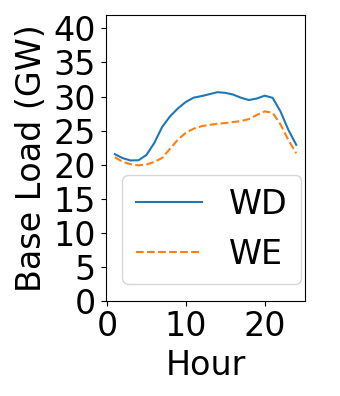}
        \caption{Spring}
    \end{subfigure}
    \begin{subfigure}{.235\columnwidth}
        \centering
        \includegraphics[scale=0.29]{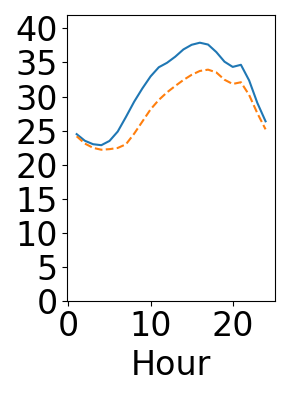}
        \caption{Summer}
    \end{subfigure}
    \begin{subfigure}{.235\columnwidth}
        \centering
        \includegraphics[scale=0.29]{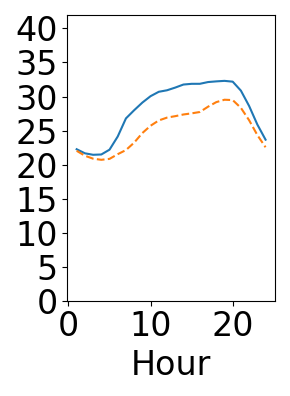}
        \caption{Fall}
    \end{subfigure}
    \begin{subfigure}{.235\columnwidth}
        \centering
        \includegraphics[scale=0.29]{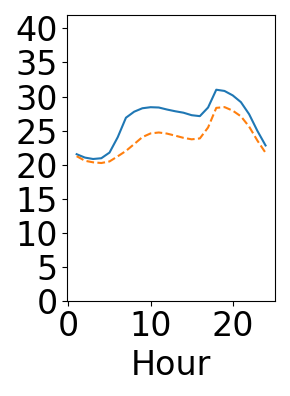}
        \caption{Winter}
    \end{subfigure}
    \caption{Grid Base (Non-DC) Load Profiles.}
    \label{fig:load}
    \Description{Daily non-datacenter load profiles for the four seasons and weekday/weekend. Given a season, weekday's and weekend's profiles have similar shapes but weekday's load is higher.}
\end{figure}

There are 8 base load, imports, and non-wind renewable generation profiles that cover the four seasons (Spring, Summer, Fall, Winter) and weekday/weekend (WD/WE). Figure \ref{fig:load} shows how each load profile varies in a day, with the average spanning from 23,780 MW (WinterWE) to 31,089 MW (SummerWD). We assume high accuracy of base load and renewable generation forecast, so the day-ahead and real-time OPF share the same deterministic load and renewable generation profiles.
%Comparing the weekday and weekend in a season, there is a clear weekly pattern that the two profiles are in similar shape but weekend's load is lower. 

\begin{figure}[h]
    \begin{subfigure}{.27\columnwidth}
        \centering
        \includegraphics[scale=0.29]{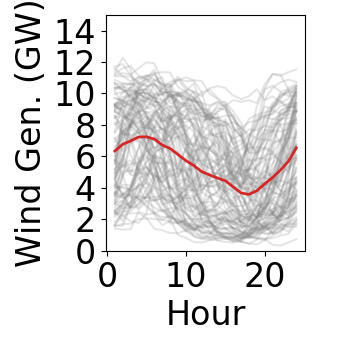}
        \caption{Spring}
    \end{subfigure}
    \begin{subfigure}{.235\columnwidth}
        \centering
        \includegraphics[scale=0.29]{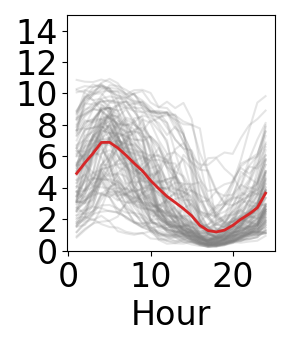}
        \caption{Summer}
    \end{subfigure}
    \begin{subfigure}{.235\columnwidth}
        \centering
        \includegraphics[scale=0.29]{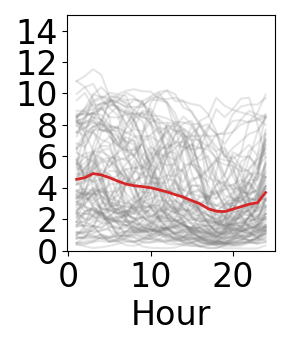}
        \caption{Fall}
    \end{subfigure}
    \begin{subfigure}{.235\columnwidth}
        \centering
        \includegraphics[scale=0.29]{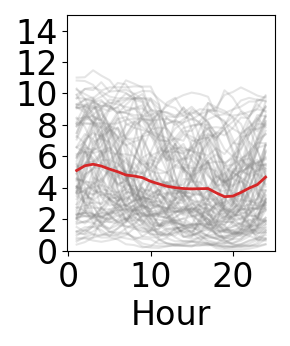}
        \caption{Winter}
    \end{subfigure}
    \caption{Used Wind Scenarios (15\% Penetration). Red lines represent average wind scenarios for each season.}
    \label{fig:windScenarios}
    \Description{There are 100 wind scenarios used for each of the four seasons. The wind generation tends to be higher in late night or early morning, which is a misalignment with load.}
\end{figure}

To reflect the impact of wind variation, for each season, we use 100 wind scenarios (Figure \ref{fig:windScenarios}) shared by the weekday and weekend. The wind generation tends to be higher in the late night and early morning, which is a misalignment with load and can be opportunities for datacenter capacity adaptation.

Datacenters are added to random buses in the grid (with loads added to the base load profiles), which reflects the fact that datacenter site selection is more based on business considerations external to the power grid (e.g. tax breaks, jobs, internet hookups, etc.).

\subsection{Evaluation Metrics}
\label{subsec:metrics}

Our evaluation metrics cover both datacenter goals and impacts on the grid and other grid customers:
\begin{itemize}
    \item \textbf{Datacenter Carbon Reduction.} We vary only datacenter capacity, and thus 
    %adaptation accounts for the grid carbon emissions change, we report the normalized 
    attribute the grid carbon reduction to DC capacity adaptation, reporting it as percentage reduction in datacenter operational carbon emissions:
    $$\frac{(gridCarbon_{fixed-cap}-gridCarbon_{adaptation})}{datacenterCarbon_{fixed-cap}}*100\%$$
    and grid carbon emissions are calculated as:
    $$gridCarbon=\sum_{t=1}^{24} gen_{f,t}*emissionRate_f$$
    where $gen_{f,t}$ is generation from fuel $f$ in the $t$-th hour. Fuel emission rates are from US EPA eGrid database \cite{eGrid, caisoGHG} and listed in Appendix \ref{sec:fuelCI}. $datacenterCarbon$ with fixed-capacity DCs is calculated using the grid emission rate. 
    %With 30 datacenters, the percentage DC carbon reduction is roughly 8 times of percentage grid carbon reduction.
    \item \textbf{Grid Dispatch Cost (\$)} is the objective for DC-OPF minimization and a figure of merit for grid operation.
    %the lower the better.
    \item \textbf{Datacenter Average Power Price (\$/MWh)} is how much the datacenter would pay for power considering the location of datacenter and the time when power is consumed.  For multiple datacenters, we compute the average. % across them.
    \item \textbf{Non-datacenter Customer Average Power Price (\$/MWh)} is the average power price across grid customers other than datacenters, weighted by power demand.
    \item \textbf{Datacenter Average Capacity Variation (MW/h)} is defined as the average change in power capacity of a datacenter between adjacent one-hour periods, the lower the better.  More formally:
    $$\frac{1}{23}\sum_{t=2}^{24}|cap_{i,t}-cap_{i,t-1}|$$
    where $cap_{i,t}$ denotes the capacity of datacenter $i$ at time $t$.
\end{itemize}
Conventional datacenters operate at fixed capacity; we use this as the baseline when examining the impacts of datacenter capacity adaptation.
%To measure how effectively datacenters with different adaptation approaches can achieve their objectives, we normalize the results to a grid-controlled optimization (\textbf{GridControl}). As this grid optimization adapts datacenter loads to time periods with surplus renewable generation to maximize social welfare, its results can be regarded as \textbf{potential} datacenter and grid benefits.

\subsection{Experiment Setup}
Given the 100 wind scenarios for each day type, our simulation is equivalent to simulating a total of 800 days \cite{papavasiliou2013multiarea}. For each day, we vary the wind penetration level and simulate datacenter operation using different capacity adaptation algorithms. The results reported in Section \ref{sec:evaluation} are the average of 8 day types (weighted for the number of weekdays and weekend days) and varied wind scenarios.

We used Julia 1.5.2 with JuMP v0.21.5 \cite{dunning2017jump} to implement the grid simulation and solved grid OPF with Gurobi Optimizer v9.0 \cite{gurobi}.

\section{Evaluating Datacenter Capacity Adaptation Approaches}
\label{sec:evaluation}
We explore datacenter capacity adaptation algorithms for 
each of the three approaches in Section \ref{sec:approach}.
Evaluation uses a full grid simulation and results are compared to datacenters with fixed-capacity (no capacity adaptation).
%The comparison is based on datacenter carbon reduction, as well as cost impacts for datacenters and other customers.
%impacts on datacenters and other customers in the grid.
%We start from comparing the effectiveness of different grid metrics in guiding datacenter power adaptation for carbon reduction, identifying LMPrice as the most effective one. Then we examine improvements for LMPrice-based load adaptation. Finally, we show capacity-plan sharing significantly outperforms them and achieves benefits other than carbon emissions. 
While we have studied many scenarios, for clarity we show a single representative
%focused on a representative 
scenario (30 200-MW datacenters and dynamic range of [0.4, 1.0]).  In this scenario,
datacenters are 10\% of grid load, less than in leading-edge
grids, but realistic for dozens of grids throughout the world in the near
future.

\subsection{No Coordination (Local Scope)}
\label{subsubsec:onlControl}

With local, online capacity adaptation, each datacenter makes
real-time capacity decisions independently based on current and future 
grid metrics.  We employ a dynamic programming algorithm that
makes hourly decisions using current value of metric and its daily
average, and refer to local adaptation based on different metrics by ``\textbf{<Metric> (Avg)}''.  
The algorithm selects amongst $\{avgCap, avgCap\pm \text{dynamic range}/2\}$ to minimize the expectation: $$cap_{i,t}*metric_{i,t} + (backlog_{i,t-1}+avgCap-cap_{i,t})*metric_i$$ which can be either carbon emissions or power cost.  %which represents the expected carbon emissions or power cost according to the used metric. 
The $backlog$ is then updated given the determined power capacity.

Datacenter capacity adaptation is coupled to grid dispatch (OPF optimization), as below, where datacenters are denoted by $i$, and hours by $t$:

\begin{enumerate}
    \item For all i,t: $cap_{i,t}=avgCap$ (neutral initial condition).
    \item Solve grid OPF with $\{cap_{i,t}\}$, defining $metric_{i,t}$ as day-ahead information. %each DC for all 24 hours.
    \item At the beginning of $t=1,...,24$-th hour, each datacenter adapts capacity $cap_{i,t}$ based on the metrics. 
    \item Then the grid solves OPF with updated datacenter capacities $\{cap_{i,t}\}$ (real-time operation).
    \item This new OPF solution redefines $metric_{i,t}$ (realized) and for the future (i.e. $[t+1,...,24]$).
\end{enumerate}

\subsubsection{Comparing Grid Metrics.}
\begin{figure}[h]
    \centering
    \includegraphics[width=\columnwidth]{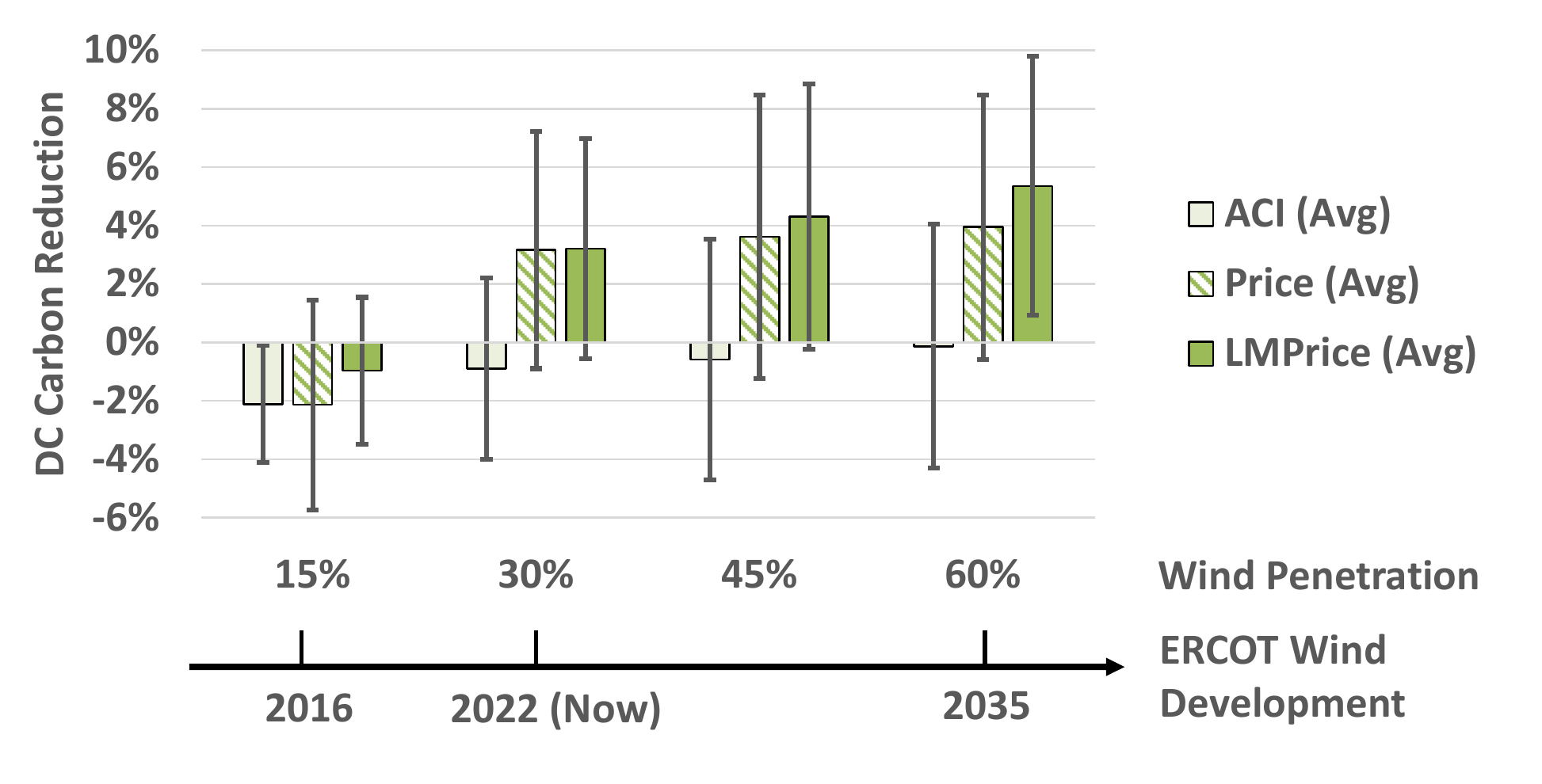}
    \caption{DC Carbon Reduction with Adaptation based on Carbon Intensity, Grid Price, and Locational-Marginal Price.}
    \label{fig:dcCarbon_metric}
    \Description{DC Carbon Reduction with Adaptation based on Carbon Intensity, Grid Price, and Locational-Marginal Price. The x-axis is 15\% to 60\% wind penetration and represents expected progress of renewable generation in ERCOT. Adaptation with Locational-Marginal Price achieves the most carbon reduction.}
\end{figure}

We evaluate the effectiveness of different grid metrics used for adaptation---average
carbon intensity (ACI), grid price (Price), and locational marginal
price (LMPrice).
LMPrice consistently outperforms
ACI and Price (Figure
\ref{fig:dcCarbon_metric}).

The x-axis in Figure \ref{fig:dcCarbon_metric} is (15\%--60\% wind penetration) and represents expected progress of renewable generation; for calibration we use the expected timing of this change for ERCOT as a reference.  We use this x-axis in many figures. We also plot standard deviation across daily variation in renewable generation (wind scenarios) as ``whiskers''.  Each specific wind scenario
typically produces correlated results for different algorithms, so
these whiskers capture variability, not uncertainty.

At 15\% wind, online adaptation fails to reduce 
carbon-emissions because 
generation supply is tight, % datacenter load adaptation can
and oversubscription happens during low-carbon periods.  Higher wind penetration
provides more opportunity, enabling online adaptation using LMPrice to reduce
datacenter carbon emissions by 5.4\% (60\% wind), consistently outperforming
ACI (+5.4\%) and Price (+1.4\%),  Our broader studies show that price metrics work better generally, and finer-grained pricing works best.

\begin{figure}[h]
    \centering
    \includegraphics[width=\columnwidth]{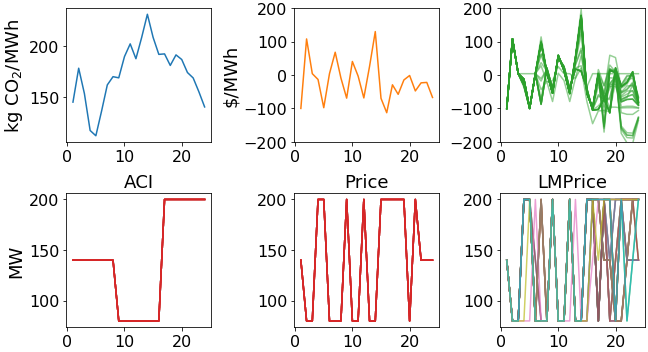}
    \caption{Global vs. Local metrics (LMPrice) create different DC capacity  adaptation (bottom).  Local produces DC adaptation variety. (Fall Weekday, 30\% Wind Penetration)}
    \label{fig:metric_dcload}
    \Description{Examples of adaptation behaviors with the three different grid metrics. Local metric (locational-marginal price) produces adaptation variety.}
\end{figure}

To illustrate, Figure \ref{fig:metric_dcload} shows a single-day
timeline.  The graphs show that the grid-wide signals, ACI and Price, create uniform, lockstep datacenter behavior, maximizing the load
swings for the grid.  
%However, as the power grid is a network with constraints
%(e.g. transmission), the impact of load adaptation is
%location-specific.
In contrast, the locational metric, LMPrice, produces diverse behaviors as each datacenter reacts to local pricing that reflects grid constraints such as transmission and generator ramping.

%%such constraints and local low-carbon periods. Overall, LMPrice guides load adaptation most effectively among the metrics we study, and its availability to market participants makes it a practically usable for datacenter power adaptation.

\subsubsection{Improving Local Online Adaptation.}
\label{subsec:sophisticatedOnline}

Using LMPrice, we next explore how to improve local-scope adaptation by adding finer-resolution price , better forecasts, 
and smoothing capacity changes.

Using the day-ahead hourly locational prices at the DC site (price array $\{\text{predLMP}_{i,t}\},t=1,...,24$), in the $j$-th hour, datacenter $i$ performs dynamic programming on the price array $\{p_{i,t}\}$ with:

\begin{equation*}
p_{i,t}=
\begin{cases}
&\text{LMP}_{i,j},\ \text{if }t=j\text{ and }j\neq 1\\
&\text{predLMP}_{i,t},\ \text{otherwise}\\
\end{cases}
,\ t=j,...,24
\end{equation*}
where $\text{LMP}_{i,j}$ is the real-time price after $\{cap_{i,j-1}\}$ are set. The dynamic programming algorithm produces a capacity array based on the following recurrence formula:
\begin{equation}
\begin{split}
    cost_i(n, t, cap)=cap*p_{i,t}+\min_{cap'}&\{cost_i(n+cap-avgCap, t-1, cap')\\
    &\ |\  |cap-cap'|\le stepSize\}
\end{split}
\end{equation}
where $cost_i(n,t,cap)$ denotes the minimum power cost of the sub-problem ending at $t$-th hour with backlog $n$ and capacity level $cap$ in the $t$-th hour. Following the convention of dynamic programming, the resulting capacity array is obtained through backtracking from $min(\{cost_i(0,24,cap)\ |\  cap_{min}\le cap\le cap_{max}\})$, which is the optimal cost subject to the constraints defined in Section \ref{subsec:flexibility}. The datacenter takes the first element of the solution array (the last in backtracking) as $cap_{i,t}$---similar to the receding horizon control \cite{kwon2006receding} but with a fixed horizon. This algorithm models:

%reduction among our collection of grid metrics, we then seek to improve LMPrice-based adaptation approaches. However, these improvements can be applied to adaptation based on other metrics with similar data availability as well. We consider the followingenhancements:

\textbf{Forecasts (Future Information).} We use hourly LMPrice from
the day-ahead market as a price forecast.  It is refined in other markets (e.g. hourly, 15-minute, 5-minute, real-time) as the time approaches until the final OPF determines the dispatch, prices, etc. Adding this information---the full 24 hours of day-ahead prices---is reflected in \textbf{LMPrice (Hourly)}.
%These forecasts are not guaranteed to be accurate as the final OPF that runs at the specific hour will determine dispatch, prices, etc.
% so that it can capture low-carbon time periods better.
%gets more information about daily price variation and can better estimate the future price curve. 

\textbf{Step Size (Smoothed Capacity).} %\aac{Need to reorg/rewrite}
Online adaptation can cause large datacenter capacity fluctuations, harming both datacenter computing efficiency and the grid (generation dispatch, carbon intensity, price). To smooth these large capacity changes, we limit hour-to-hour DC capacity change with a maximum step size.

\begin{figure}[h]
    \centering
    \includegraphics[width=\columnwidth]{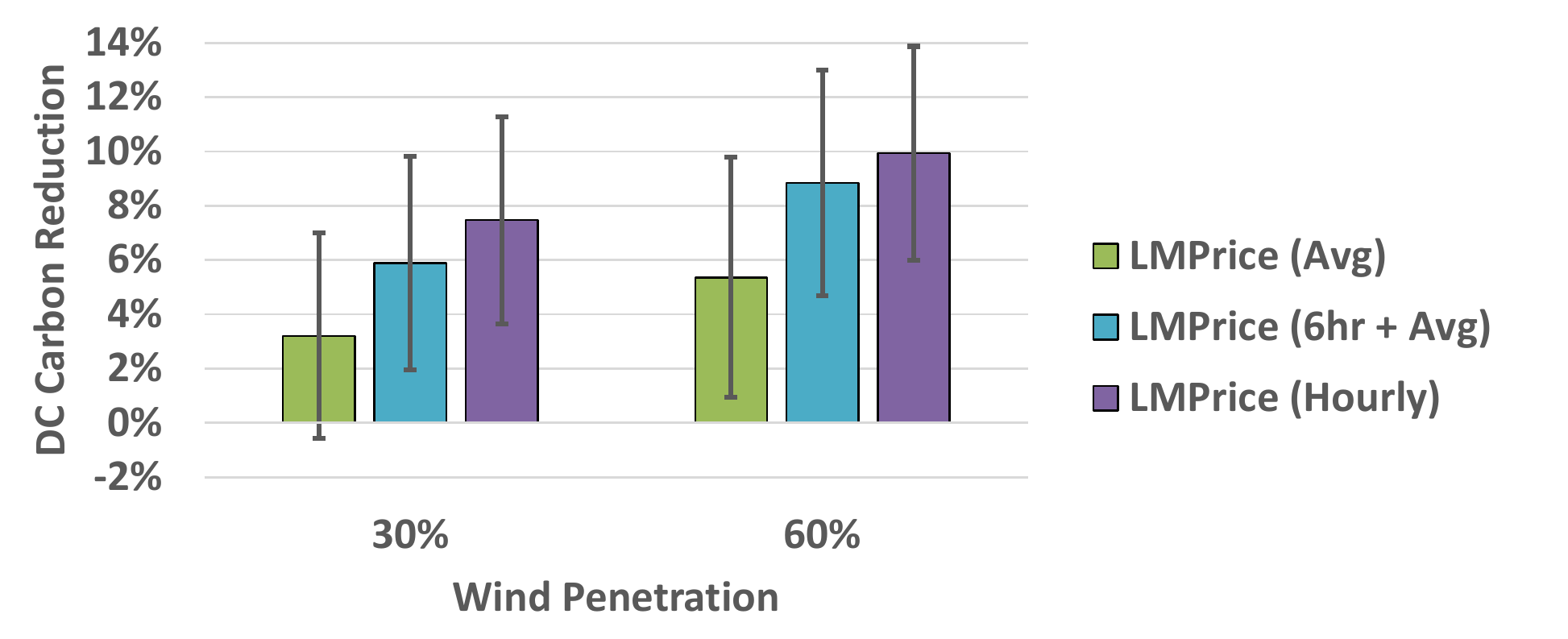}
    \caption{Datacenter Carbon Reduction with varied future LMPrice information  (Day average, 6hr detailed, 24hr detailed).}
    \label{fig:dcCarbon_onl_hour}
    \Description{Datacenter Carbon Reduction with varied future LMPrice information. The performance improves as future LMPrice information gets more detailed}
\end{figure}

Using detailed future price information, we empirically determined that a 
step size of 40 MW/h for LMPrice (Hourly), and unbounded for LMPrice (Avg) that yield the largest carbon reduction.  50\% of the improvement from LMPrice (Avg) to LMPrice (Hourly) is due to price information and 50\% to step size (Figure \ref{fig:dcCarbon_onl_hour}).

%\aac{maybe put 6hr in here}
%We first tune the step size for AdaptLMPrice-Avg and AdaptLMPrice-Hourly, limiting the power level change from hour to

%Figure \ref{fig:dcCarbon_onl_hour}.  The results reflect the best step sizes (40 MW/h for AdaptLMPrice-Hourly, and no limit for AdaptLMPrice-Avg). Our drill-down analysis finds that AdaptLMPrice-Hourly's improvement is roughly 50\% from hourly information and 50\% from step size.

%To further study the impact of detailed future information, we develop a variant, which at each hour is given hourly future information for the next 6 hours but only daily average after that. The two approaches with detailed future information are clearly superior: the variant given 6-hour future price information (\textbf{AdaptLMPrice-6hr+Avg}) already outperforms \textbf{AdaptLMPrice-Avg}, and benefit grows with more future information in \textbf{AdaptLMPrice-Hourly}.  

% , which uses hourly information for the following 6 hours but daily average for rest of the day. Clearly, hourly future information improves the datacenter carbon reduction results.

In Figure \ref{fig:dcCarbon_onl_hour}, \textbf{LMPrice (6hr + Avg)} reflects progressively introduced future information, which is given hourly information for the next 6 hours but only daily average after that. For 30\% wind penetration, 6 hours'
information enables 2.6\% reduction and 24 hours'
a 4.2\% reduction in datacenter carbon emissions. These 
grow to 3.5\% and 4.6\% respectively at 60\% wind penetration.  Ultimately  LMPrice (Hourly) delivers 10\% of DC carbon reduction.

%We consider two local datacenter load adaptation algorithms, Indep-ONL-Avg that uses a local, average price for the whole day, and Indep-ONL-Hourly that uses hourly estimated future prices.
%This second algorithm can account for expected future price fluctuations (e.g. increase evening load). We explore how these algorithms affect DC renewable fraction and power price.

%in power price to achieve lower cost and higher grid benefit (low price usually indicates excess renewables or import curtailment). 

%, at the cost of a more fluctuating load (capacity) curve.

%%However, it also potentially produces more fluctuating capacity curve, which can harm the grid and datacenters.

%%The harm mitigates as wind penetration increases, slightly outperforming Indep-ONL-Avg in grid benefit at 60\% wind.

%Results with different step sizes form the lines in Figure \ref{fig:dcCarbon_onl_hour_step}. Under the same wind penetration level, the DC RPS first tends to increases as the step size smooths the DC load (step size$\ge$40 MW), reaching about 1.3x improvement (+0.2\% DC RPS) vs. Indep-Ahead-Hourly with 40 MW/h step size. As the step size continues to decrease, the DC RPS slightly decreases, suggesting limited load flexibility can limit the achieved benefits. DC power cost is reduced as well: with 40 MW/h step size, Indep-ONL-Hourly-Step produces significantly lower cost than Indep-ONL-Avg, but the cost is still slightly higher than the fixed DC load at 15\% or 30\% wind penetration.

\subsection{Coordinating a Group of Datacenters}
\label{subsec:coordination}

Independently controlled DCs can react together, when decisions are based
on grid-wide or other strongly-correlated metrics,
producing a large aggregate power capacity change.  We have seen LMPrice is better, but even its
correlation across sites can produces synchronized DC capacity
changes that are difficult for the grid to manage.  Addressing this,
we add an external limiter, called ``coordinator'', to mitigate the harm.
In \textbf{LMPrice (Avg)-Coord}, each datacenter runs independent
online control algorithm and then submits their adaptation to a
coordinator.  The coordinator limits total power change for a set of
datacenters, using a quota:
\begin{enumerate}
    \item Generate a random permutation of datacenters.
    \item For each datacenter, if $change \le quota$ then accept, $quota = quota - change$; else, reject.
    %\item reject the rest of the changes
\end{enumerate}

If local controllers were unable to get their requested change, they must  update their backlogs accordingly.

%Each DC runs its own independent online control algorithm, but then submits the decision to a coordinator.  The coordinator limits the total power change of a group of DCs (up or down) with a fixed quota; denying some DC changes.

\begin{figure}[h]
    \centering
    \includegraphics[width=\columnwidth]{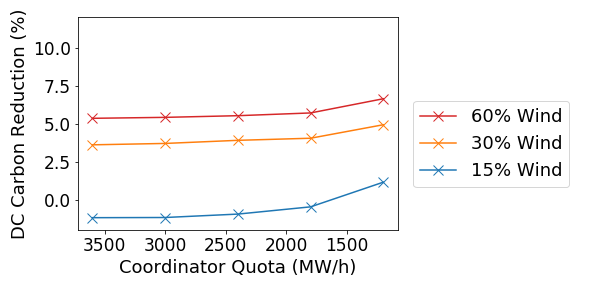}
    \caption{Datacenter Carbon Reduction vs. Coordinator Quota for three wind penetration levels.}
    \label{fig:dcCarbon_quota}
    \Description{DC carbon reduction with 3600 to 1200 MW/h coordinator quotas.  Each line represents a different wind penetration level. Coordination improves performance, mitigating overshifting harm at 15-60\% wind penetration: on each line, all the points reflect greater carbon reduction than LMPrice (Avg), and steady improvement as the quota is tightened.}
\end{figure}

Figure \ref{fig:dcCarbon_quota} shows DC carbon reduction
with varied coordinator quotas (3600--1200 MW/h).  Each line
represents a different wind penetration level.
Coordination improves performance, mitigating overshifting harm at 15-60\% wind penetration:
on each line, all the points reflect greater
carbon reduction than the leftmost one---LMPrice (Avg), and steady improvement as the quota is tightened.
The benefit for datacenters is larger when the overshifting is more evident under tight generation (15\% wind penetration), improving DC carbon reduction by
2.3\%.  These benefits are smaller than those achieved by exploiting future
price information, such as LMPrice (Hourly), which yields a 50\% higher DC carbon reduction.

%%results (about 66\% of AdaptLMPrice-Hourly).

\paragraph{Multiple Coordinators.}
% move some text to methods
In many geographic areas, there are multiple cloud providers (e.g. Northern Virginia, Texas, Ireland, or China's Ningxia), and each cloud provider has multiple datacenter sites in that area. As competitors, they may not be willing to share a coordinator. To model this multi-coordinator scenario, we assume there are 2 or 3 coordinators in the grid, each coordinating 10 or 15 DCs. 
We use an overall quota of 1200 MW/h, and divide it equally across the coordinators.   

%%which achieves relatively improvements in dispatch cost with little unmet demand, and evenly distribute this quota to the coordinators. 

Increasing the number of coordinators further decreases DC carbon emissions. The reason for this is narrower coordinator scope with datacenter load quantization increases the smoothness of total datacenter load, but the improvements are small. 3 coordinators produce improvements up to 0.7\% of DC carbon emissions, achieving 7.3\% of DC carbon reduction at 60\% wind penetration.

%%\textit{Summary.} Coordination that smooths the aggregate DC load change mitigates overshifting, but it's still not effective enough to solve the problem.

\subsection{Coordination with the Grid: Capacity Plan Sharing}
\label{subsec:loadPlanSharing}

Adaptive datacenters cause grid problems as their large power changes are unpredictable and strain generator ramp constraints.  Further, unplanned adaptation can produce rapid changes in compute capacity, making it difficult for cloud resource managers to be efficient.
In view of these insights, we expand the space scope to datacenter-grid cooperation and time scope to day-ahead (planning), proposing a new approach---\textbf{PlanShare}: datacenters create a 24-hour adapted capacity plan based on LMPrice in day-ahead grid market \cite{caisoMarketProcess} ahead of operating day, and then share the plan with the grid. This allows the grid to optimize globally based on the DC information. Formally, with datacenters denoted by $i$ and hours denoted by $t$:

\begin{enumerate}
    \item For all i,t: $cap_{i,t}=avgCap$ (neutral initial condition).
    \item Solve grid OPF with $\{cap_{i,t}\}$, defining initial $\text{LMP}_{i,t}$ as day-ahead information.
    \item Each DC makes 24-hour adaptation plan using LMPrice (Hourly)'s dynamic programming algorithm and shares it with the grid.
    \item Solve grid OPF for $[1,...,24]$ with adapted $\{cap_{i,t}\}$ to model the next day's operation---the datacenter must follow the full-day capacity plan it shares with the power grid.
\end{enumerate}

\begin{figure}[h]
    \centering
    \includegraphics[width=\columnwidth]{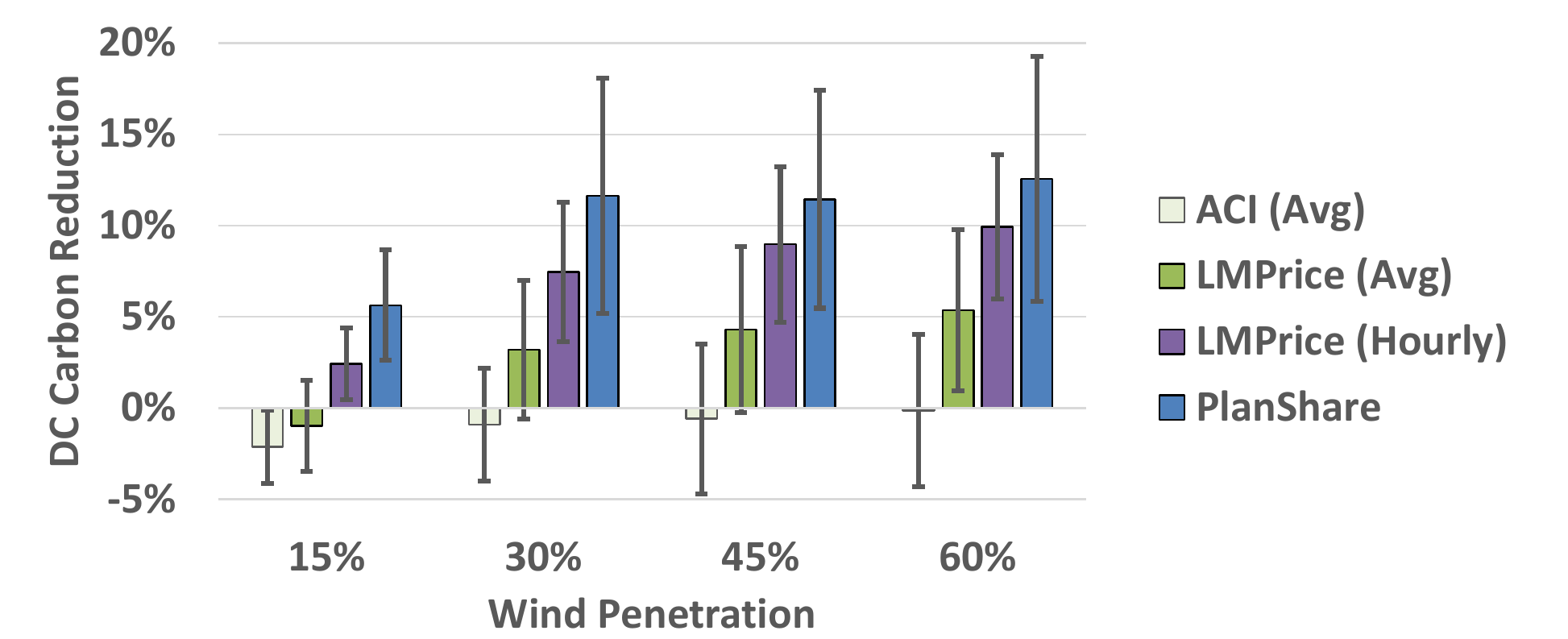}
    \caption{DC Carbon Reduction for four capacity adaptation schemes.  PlanShare outperforms all Online approaches and under all wind penetration levels.}
    \label{fig:dcCarbon_planshare}
    \Description{DC Carbon Reduction for four capacity adaptation schemes.  PlanShare outperforms all Online approaches and under all wind penetration levels, up to 12.6\% carbon reduction at 60\% wind penetration.}
\end{figure}

\begin{figure}[h]
    \centering
    \includegraphics[width=\columnwidth]{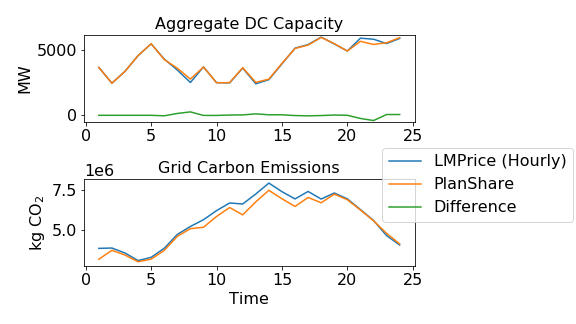}
    \caption{PlanShare produces similar aggregate DC capacity (Top)  %to AdaptLMPrice-Hourly, 
    but sharing information with the grid produces 8x larger DC emissions reduction (Bottom). (Fall WD Example)}
    \label{fig:metric_dcload_planshare}
    \Description{Compared with LMPrice (Hourly), PlanShare produces similar aggregate DC capacity, but sharing information with the grid produces 8x larger DC emissions reduction}
\end{figure}

Figure \ref{fig:dcCarbon_planshare} compares PlanShare
and the local, online approaches.  The results show the benefits of plan sharing
with the grid.  Using essentially the same adaptation algorithm (comparison of behaviors shown in Figure \ref{fig:metric_dcload_planshare}), and on less accurate information,
by working with the grid, PlanShare reduces DC carbon emissions by
up to 12.6\% (1.6\% grid carbon reduction if normalized to grid carbon emissions).  This is 1.26x better than the best local, online adaptation
result, and the advantage is even higher (1.56x) with today's wind
penetration levels (30\%).  By contributing its adaptation plan to
the grid optimization---in advance and as a committed schedule---PlanShare dramatically improves the datacenter carbon emissions reduction
that can be achieved.

%in future DC loads in turn improves the realized benefits of DC power adaptation!

\paragraph{Sensitivity to Length of Shared Plan.}

\begin{figure}[h]
    \centering
    \includegraphics[width=\columnwidth]{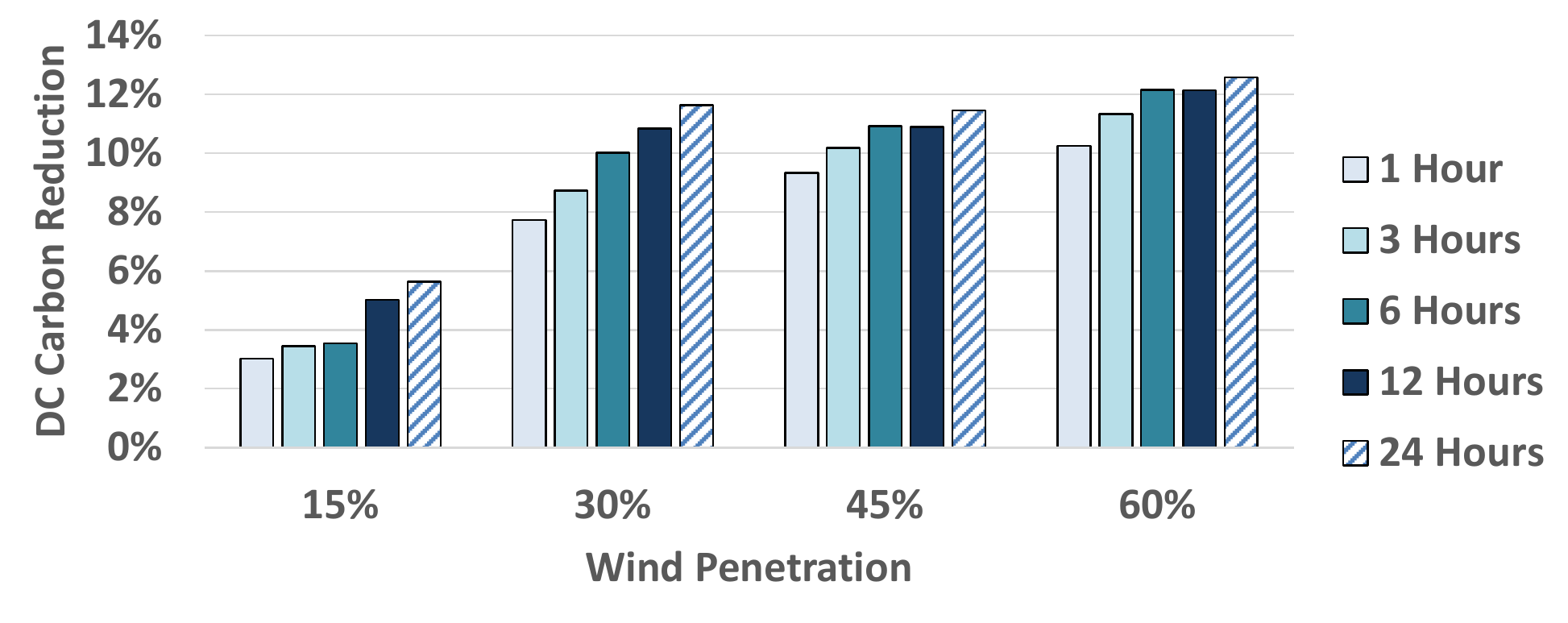}
    \caption{DC Carbon Reduction vs. Varying Plan Length: PlanShare with 1-24hr plan.}%that share from 24 hours down to only one hour of planned loads.}
    \label{fig:dcCarbon_ahead_len}
    \Description{Sensitivity of PlanShare carbon reduction to length of shared plan that ranges from 1 to 24 hours. A single hour advance plan is enough for PlanShare to match the performance of LMPrice (Hourly), and the performance improves significantly as length is increased.}
\end{figure}

Having demonstrated the benefits of a practical scheme (24-hour
day-ahead plans are available in many power grids),
 we explore {\bf how much DC plan information is needed by the grid?}  We vary the length of the DC capacity plan shared
from 1 to 24 hours, reporting results in
Figure \ref{fig:dcCarbon_ahead_len}.  Interestingly, a single hour advance plan
is enough for PlanShare to match the performance of LMPrice (Hourly), the best
online approach, despite the fact that PlanShare has no online adaptation; the full 24-hour capacity plan is fixed.
%%
%% This raises interesting questions as to the value of online techniques. [Note this Peter!!!!!]
%%
As the length of plan is increased to 3, 6, and 12 hours, the benefit of plan sharing increases significantly, reaching the large benefits previously highlighted in Figure \ref{fig:dcCarbon_planshare}.

%% clearly shows grid optimization with more future DC load information improves the effectiveness of DC load adaptation.

% To study how sensitive the grid optimization is to future datacenter load information, we consider sharing shorter plans with the grid. we keep the same day-ahead adapted power plan but vary the time length of and frequency of sharing schedule from every hour, every 3 hours to sharing whole day's schedule at a time (i.e. PlanShare).

\subsection{Datacenter Adaptation Impacts beyond Carbon}
\label{subsec:otherDCImpact}
Datacenter capacity adaption produces other impacts on the grid customers and datacenter operation.  We study
several here.

\begin{figure}[h]
    \centering
    \includegraphics[width=\columnwidth]{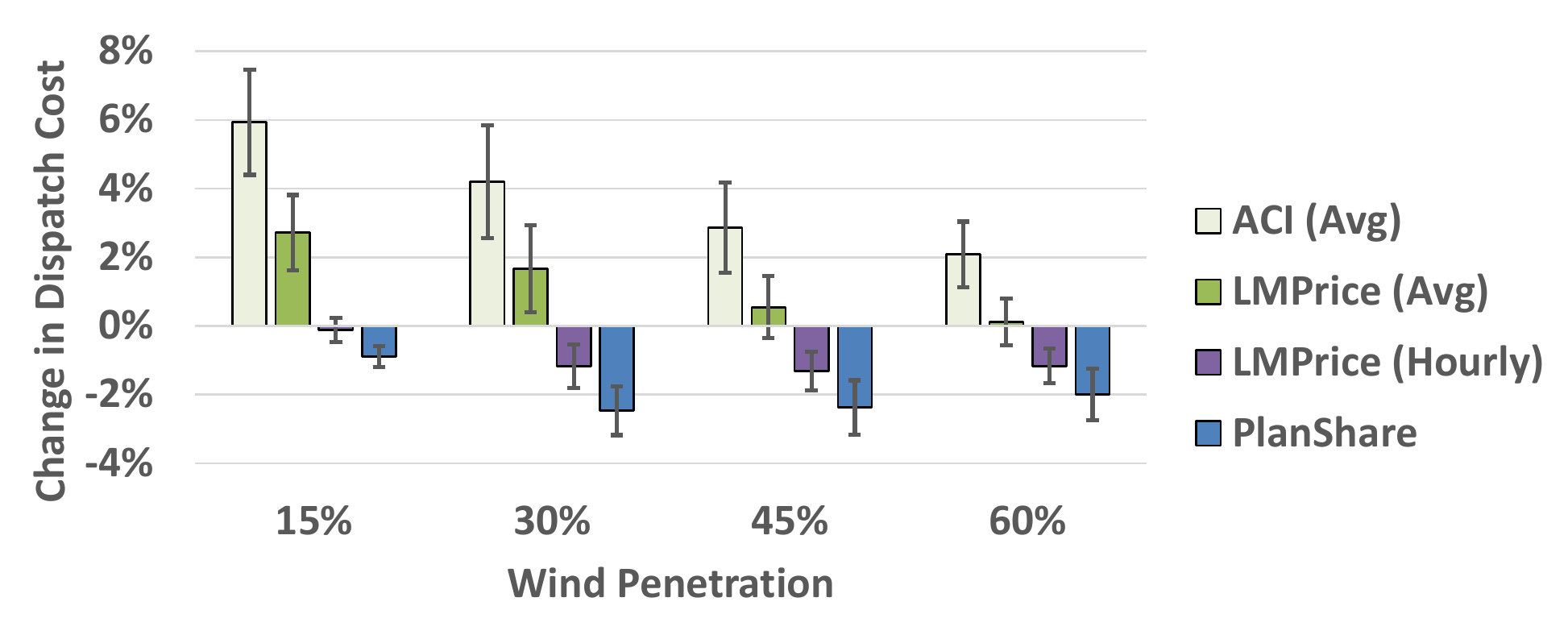}
    \caption{Grid Dispatch Cost vs. Adaptation Scheme---PlanShare reduces grid dispatch cost most.}
    \label{fig:dispatchCost_summary}
    \Description{Grid dispatch cost for four adaptation schemes. PlanShare reduces grid dispatch cost most.}
\end{figure}

\paragraph{Grid Dispatch Cost.}
We compare the grid dispatch cost of different approaches in Figure \ref{fig:dispatchCost_summary}. Local, online datacenter adaptation approaches can increase grid dispatch cost at lower wind penetration (up to 6\% increase by ACI (Avg)).  PlanShare successfully eliminates this grid performance damage, decreasing grid dispatch cost by as much as 2.5\%. The reduction is attributed to decreased generation cost and renewable curtailment penalties. Datacenter capacity adaptation and grid-wide optimization given the shared plans enable this.

\paragraph{Datacenter Power Cost.}

\begin{figure}[h]
%\textbf{Should we split this into two figures?}
%\textbf{Perhaps not as spliting (and expanding each to 15\% to 60\% wind) doesn't add much new information}
    \begin{subfigure}{.199\textwidth}
        \centering
        \includegraphics[width=\columnwidth]{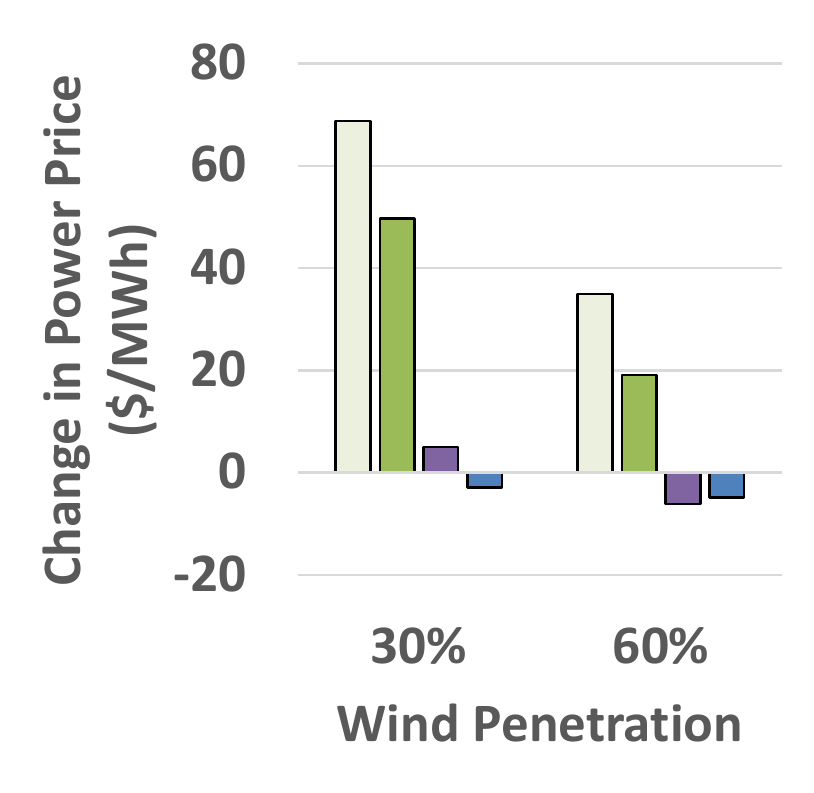}
        \caption{Datacenters.}
    \end{subfigure}
    \begin{subfigure}{.27\textwidth}
        \centering
        \includegraphics[width=\columnwidth]{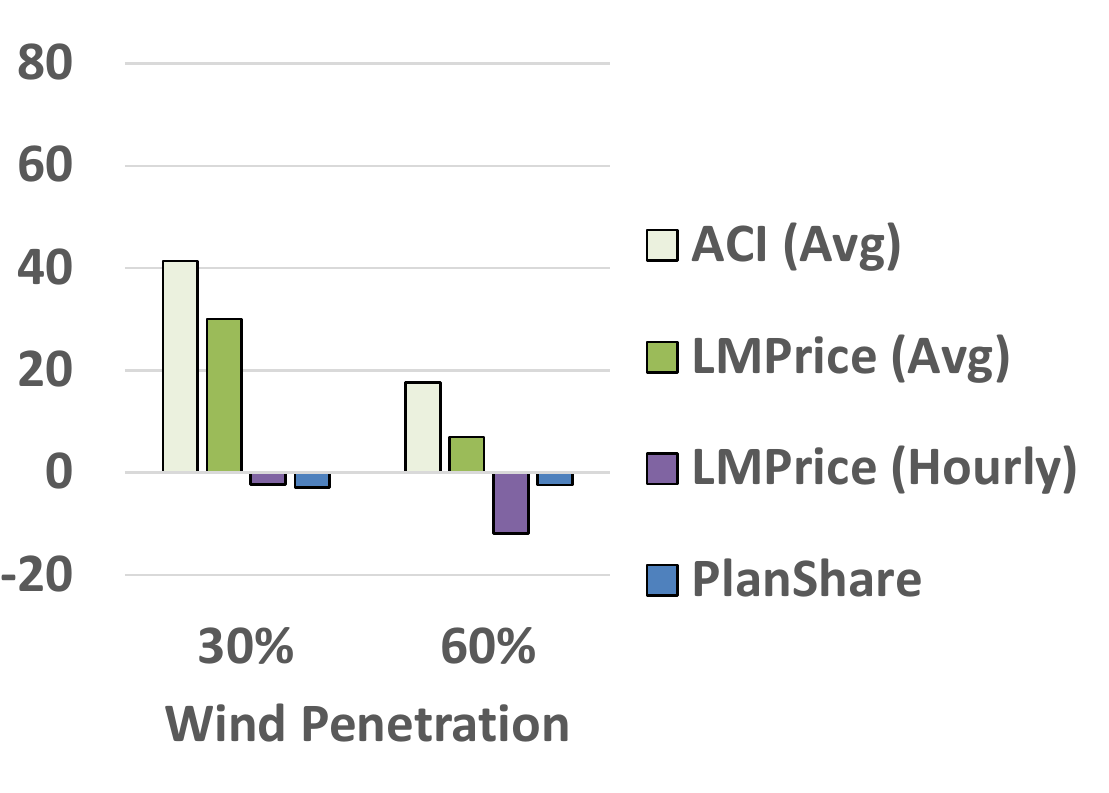}
        \caption{Other (Non-DC) Customers.}
    \end{subfigure}
    \caption{Change in Average Power Price for Different Grid Customers.}
    \label{fig:price}
    \Description{Change in average power price for datacenters and other grid customers. PlanShare can decrease datacenter power price by up to 30\%, while stably benefiting other customers as well.}
\end{figure}

Adaptive datacenters affect power pricing in the grid.  In Figure
\ref{fig:price}a, results show that local, online adaptation
can cause significant power prices increases of \$20 to \$50/MWh,
corresponding to 59\%--490\% increase in power cost.  This is because
locally controlled adaptation clashes with grid constraints and
dynamics (overshifting in Section \ref{sec:problem}), and it is likely a
major deterrent for datacenter adoption.  In contrast,
sharing the datacenter's capacity plan in advance as in
PlanShare decreases average power price stably, up to 30\% compared
with the fixed-capacity scenario.

\paragraph{Non-DC Customer Power Cost.}
%Adaptive datacenter power-level management to reduce carbon emissions shouldn't make other grid customers to have higher prices.  

In Figure \ref{fig:price}b, We explore how adapting
datacenter capacity impacts other (non-DC) customers.  Local approaches,
including ACI (Avg) and LMPrice (Avg), significantly increase
the price for non-DC customers, especially for lower wind penetration
with less excess renewable generation (LMPrice (Hourly) also
increases price at 15\% wind).  Datacenters as growing consumers of power
are already subject to growing scrutiny and negative publicity.  Pricing
hard to other customers has the potential to cause a
backlash, so datacenters should be careful about deploying such local,
online adaptive capacity control. %AdaptLMPrice-Hourly and
PlanShare avoids this price harm for non-DC customers at all wind
penetration levels.
%Increased wind penetration decreases competition for power, and thus prices for all.
%Capacity plan sharing also significantly reduces the harm on the power grid. At 15\% wind penetration, Indep-Ahead-Hourly increases dispatch cost by only 1.3\%, much lower than Indep-ONL-Avg (2.7\%) and Indep-ONL-Hourly (4.4\%), and it reduces dispatch cost by 1\% at 60\% wind.

%\aac{move to end of section 6?}
%Worse still, from the grid perspective, it can produce up to 4.4\% higher dispatch cost for the entire grid (approximately \$2M/day)!
%\aac{More to end section 6?}
%Similar trend is seen on grid power price---Indep-ONL-Hourly increases non-DC customers' power cost as well.
%\aac{shift to later?}
%It also decreases the grid average power price. As for grid dispatch cost and RPS, it can eliminate the harm and realize most of the potential benefits (shown later).

\begin{figure}[h]
    \centering
    \includegraphics[width=\columnwidth]{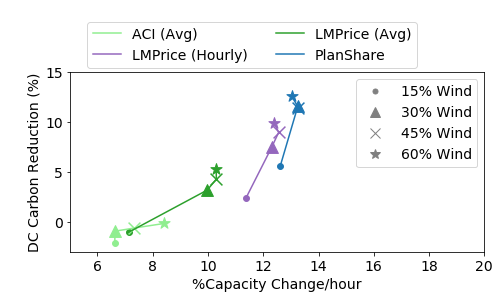}
    \caption{DC Carbon Reduction and Capacity Variation.}
    \label{fig:dcCarbon_capacityVar}
    \Description{DC Carbon Reduction and Capacity Variation. The unit of capacity variation is percentage of 200 MW capacity. PlanShare requires greater capacity variation for higher carbon reduction. However, different from others, PlanShare fixes the capacity plan in advance so the resource manager has a statically known resource schedule at the start of the day.}
\end{figure}

\paragraph{Datacenter Capacity Variation.}
Capacity variation is a critical concern for datacenter operators as
it affects workload efficiency of the available capacity.
%We first take datacenter capacity variation into consideration, the
%datacenter efficiency cost of load adaptation benefits.
Using the average capacity variation metric (see Section \ref{sec:methods}),
Figure \ref{fig:dcCarbon_capacityVar} shows DC carbon reduction on y-axis and
capacity variation on x-axis. As before, DC carbon reduction is
relative to the fixed-capacity scenario, and capacity variation (MW/h)
is normalized to the maximum capacity (200 MW). An ideal adaptation
approach would fall in the upper-left corner (high carbon reduction,
low capacity variation). Each line connnects results as wind penetration
increases for a given adaptation approach.  

\begin{table*}[htb]
\caption{Summary of Datacenter Capacity Adaptation Approaches}
\label{tab:summary}
\begin{center}
\small
\begin{tabular}{|p{1.25in}|c|c|c|c|c|}
    \hline
    &ACI  & LMPrice & LMPrice & PlanShare & PlanShare \\ 
     &   (Avg)     & (Avg) & (Hourly) & (1 hour)  & (24 hours) \\ \hline 
    \hline
    DC Carbon Reduction  & -- & neutral & + & + & ++  \\
    \hline
    DC Power Cost & -- -- & -- & neutral & + & ++ \\
    \hline
    DC Capacity Variation & + & + & neutral & neutral & neutral \\
    \hline
    Grid Dispatch Cost  & -- & neutral & + & + & ++  \\ 
    \hline
    %Non-DC
    Other Customers (non-DC) Power Cost & -- -- & -- & neutral & + & ++ \\
    \hline
\end{tabular}

\end{center}
\end{table*}

The plot clearly shows how the higher-performing adaptation techniques
exploit increased changes to adapt to the changing grid carbon properties.
The progression from  ACI (Avg) to LMPrice (Avg) to LMPrice (Hourly)
shows a tradeoff of carbon reduction for online capacity variation.
Achieving the greatest carbon reduction, PlanShare does
require greater capacity variation.  However, 
it's worth noting that PlanShare fixes the capacity plan in advance,
so the resource manager has a statically known resource schedule at the start of the
day, facilitating compute workload scheduling.  Scheduling studies
and other proposals \cite{zhang2021scheduling,radovanovic2021carbon, Zhang2022thesis} argue for the benefits of known capacity information. For example, with 24-hour capacity information, the cloud workloads see little performance (e.g. goodput, job wait time) degradation at 60\% dynamic range compared with the fixed capacity scenario\cite{Zhang2022thesis}.

%% All adaptation approaches see higher capacity variation as wind
%% penetration increases, because increased wind penetration increases
%% guiding metrics variation. AdaptCarbon-Avg produces the least capacity
%% variation (13--17 MW/h) but fails to reduce DC carbon emissions;
%% PlanShare reduces the much more (>1.26x) carbon emissions with
%% slightly higher (<1.1x) capacity variation than online adaptation
%% approaches.

\subsection{Summary}
Table \ref{tab:summary} summarizes the impacts of datacenter capacity adaptation in different metrics, where ``+'' means advantage and ``--'' means disadvantage. 
The weakest performance plan sharing approach---PlanShare with 1 hour's shared plan---matches and outperforms all of the other approaches.  And, as we can see in Figure \ref{fig:dcCarbon_ahead_len}, PlanShare with 24-hour adaptation plan outperforms others significantly and is by far the best, delivering the greatest DC carbon reduction while enabling datacenter resource managers to have known capacity plans. To conclude, PlanShare satisfies both datacenter (carbon and cost reduction, computing efficiency) and power grid objectives (cost reduction and fairness).

\section{Discussion and Related Work}
\label{sec:related}

%datacenter load shifting without considering the grid model (chase the sun, shift into day, shift to night, some involve storage).  focus on computing part, and there's no grid modeling.

%---information used (simple vs. some grid information)---

%selfish local control for power cost, carbon without grid coupling, online control stuff, ignores internals of DC.  could also include storage.

To the best of our knowledge, this is the first paper that explores the coordination scope for datacenter capacity adaptation and proposes sharing datacenter capacity plans with the grid to reduce operational carbon emissions. Given LMPrice metric availability, and cloud practice making day-ahead adapted capacity plan \cite{radovanovic2021carbon}, our PlanShare approach is feasible.  Some cloud providers may be reluctant to share their daily capacity plan with the grid for proprietary reasons.  But it's worth pointing out that each datacenter's local grid (e.g. Dominion Energy, PGE) already knows DC's historical power consumption, going back days, months, and years.  Further, if the cloud DCs wanted to intentionally mask their  compute load, they could still do so with on-site batteries or even generators. 

We review related work below:
\paragraph{Datacenter Capacity Adaptation (Shaping).}
Early ideas like ``follow-the-moon''``chase-the-wind'' propose to shift datacenter workload to a time or place with low energy prices or carbon emissions \cite{FollowSWM09}, exploiting variations of grid dynamics.  The DC capacity adaptation explored in this paper addresses a widely studied subset of these ideas---temporal load shaping or shifting, which typically employ sophisticated online control or optimization techniques \cite{liu2012renewable,lin2012dynamic,dou2017carbon,luo2013temporal}.  An additional variant is to manage colocated energy storage \cite{urgaonkar2011optimal,shi2016leveraging,goiri2013parasol,Lin19, Rosenthal19}. This work assumes datacenters are small loads (grid trace-based studies or DC-only evaluation), and doesn't model the impact of DC dynamic capacity changes on the grid, not to mention coupled impacts on carbon, prices, generation, etc. In this paper, a collection of hyperscale datacenters (200 MW each) coupled to a power grid model is essential to capturing their direct impacts on grid dispatch.

Recent cloud industry goals include ``24$\times$7'' or ``100/100/0'' hourly matching of DC power consumption with carbon-free generation  \cite{GoogleWhite18, msftSustainability2021}.  These efforts explore application and resource-management load shifting based on varying power carbon-intensity\cite{radovanovic2021carbon},   do not assess impact on power grids.  For example, Google's {\it carbon-aware computing} \cite{radovanovic2021carbon}, creates
day-ahead capacity plan (called a {\it virtual capacity curves} (VCC)) to enable efficient DC resource management.  But the VCC is not shared with the 
power grid.
%of upcoming capacity changes.  This perhaps could be attributed to the fact that only small percentages of capacity are under control of these schemes today.

Another dimension of load shifting is spatial or geographic, with similar goals \cite{zhou2016carbon, zhou2014fuel, Kien10, Maggs09, Sitaraman-Solar18, Liu-Greening15, Chien-Joule20, zhang2012electricity}. This  promising direction is beyond the scope of our paper.  %but it is certainly of promise.

%couple datacenter scheduling and demand response without considering datacenter impact on the grid
\paragraph{Datacenters in Demand Response.}
Datacenters' ability to delay workload (e.g. defer batch jobs) can be used to participate in demand-response programs, reducing capacity during peak periods according to requests from the grid. Such participation can reduce DC power cost, and research has explored how to balance this benefit while respecting service-level objectives (SLO) \cite{liu2013data, le2016joint}.  Further efforts in this area design sophisticated markets that incentivize DC operators and even their colocation tenants to participate in demand response \cite{chen2015greening,zhang2015truthful,sun2016online, chen2019online, zhou2018truthful}.

Demand response is designed for emergency reduction in load to protect grid stability.  As a result, actions are rare, and the power reduction is small relative to the total power consumption \cite{pjmDR2021}. In contrast, we consider multiple datacenters' active, continuous capacity adaptation for reducing operational carbon emissions with dynamic range up to 60\% of capacity. 
%as well as its impacts on both the datacenters and grid.

%studies that couple grid modeling and datacenter adaptation
\paragraph{Grid-coupled Datacenter Adaptation.}
As early as 2015, Yang, Wolski, and Chien proposed {\it Zero-carbon Cloud}, a novel approach that  proposed building datacenters as dispatchable loads controlled by the grid to harness excess carbon-free power \cite{chien2015zero,YangChien-IPDPS16,YangChien-TPDS17}, and the grid benefits including decreased dispatch cost and renewable curtailment are reported in a study including a grid model \cite{KYZC2016}. Commercial efforts at gigawatt-scale based on Zero-carbon Cloud ideas are under construction in ERCOT/Texas \cite{Lancium}. 
 Other efforts include exploring the grid benefits of temporal or spatial load shifting \cite{zhang2020flexibility,lindberg2020environmental,menati2023modeling}. \cite{lindberg2021guide} studies what grid metrics can effectively guide carbon-aware spatial load shifting. While also showing locational marginal price is effective, they claim locational mariginal carbon emissions is better, which, however, is not broadly available today.

This paper builds on the insights of \cite{lin2021evaluating}. The authors showed that without modeling the grid's dynamics, the carbon emission projections could be significantly wrong.  Thus, it's necessary to include grid models in studies of large-scale temporal workload shifting. However, that work provides no solution to coordinated management of datacenters and power grids.

%to effectively reduce datacenter's operational carbon emissions while reconciling datacenter and grid optimization, the problem focus of this paper.

%nobody is doing information sharing with the grid.  to the best of our knowledge.  here's what google and microsoft are doing... going from offset (long-term purchase) to 24x7 or 100/100/0 real-time matching.  but no information sharing -- because of their competitive concerns.

%%and may make day-ahead capacity plan \cite{radovanovic2021carbon}.   This is not shared with the grid due to proprietary concerns. 

\section{Summary and Future Work}
\label{sec:summary}
Cloud providers are adapting datacenter capacity to renewable generation to reduce  carbon emissions. However, for today's large cloud datacenters, the numerous prior techniques based on independent, online control fail to reduce emissions and can harm the grid.  To find a robust  solution, we explore the coordination of  adaptation, varying scopes in time and space.  With the local coordination scope, locational marginal price (LMPrice) is identified as a widely available and the most effective grid metric for datacenter capacity adaptation.  Expanding the scope to grid-wide coordination and day-ahead planning, we propose a solution---PlanShare, where each datacenter creates a capacity plan based on day-ahead grid metric, and then shares it with the grid.  This approach enables DCs to achieve greater emissions reduction (12.6\%), lower average power prices (-30\%), and more predictable capacity from planning (and thus better internal utilization).  PlanShare even eliminates harm to other customers in the grid.  We are optimistic that as power grids are increasingly dominated by renewable generation the techniques studied here will enable datacenters to balance the efficient delivery of cloud computing with helping balance and decarbonize the power grids.  As datacenters grow---fueled by artificial intelligence, pervasive intelligent control, commerce, and entertainment---beyond 10 and 20 percent of power grid load, these techniques will be essential not only for the power grid, but for the continued growth of computing.

Several exciting future directions for research include:  First, how to create compute load flexibility and respond to capacity change, while respecting service-level objectives (SLO)?  The resource managers and applications today lack a clear cost metric, and further it's unclear what types and extents of flexibility are possible or valuable.  Second, we have focused on temporal capacity adaptation, but it's also interesting to explore spatial shifting combined with day-ahead capacity plan schedules.  Third, how can we balance datacenter privacy considerations with the clear benefit of sharing capacity plan information?  Finally, as the world progresses to higher levels of renewable generation and computing power consumption, it will be essential to reconsider the techniques proposed here.

%%  Given the improvement from load plan sharing, a natural extension is to design load sharing mechanism that satisfies both the grid's and datacenters' requirements (e.g. flexibility, privacy). Other interesting future directions may center on how to create and incentivize load flexibility in datacenters, such as making applications and resource management carbon-aware.

\begin{acks}
We thank the anonymous reviewers for the insightful reviews, including who reviewed the earlier versions of this paper. This work is supported in part by NSF Grants CMMI-1832230, OAC-2019506, and the VMware University Research Fund. We also thank the Large-scale Sustainable Systems Group members for their support of this work!
\end{acks}

%\clearpage
\bibliographystyle{ACM-Reference-Format}
\bibliography{Bib/dcSim, Bib/zccloud-10-2020}


\begin{thebibliography}{102}

%%% the Web address.

\ifx \showCODEN    \undefined \def \showCODEN     #1{\unskip}     \fi
\ifx \showDOI      \undefined \def \showDOI       #1{#1}\fi
\ifx \showISBNx    \undefined \def \showISBNx     #1{\unskip}     \fi
\ifx \showISBNxiii \undefined \def \showISBNxiii  #1{\unskip}     \fi
\ifx \showISSN     \undefined \def \showISSN      #1{\unskip}     \fi
\ifx \showLCCN     \undefined \def \showLCCN      #1{\unskip}     \fi
\ifx \shownote     \undefined \def \shownote      #1{#1}          \fi
\ifx \showarticletitle \undefined \def \showarticletitle #1{#1}   \fi
\ifx \showURL      \undefined \def \showURL       {\relax}        \fi
% invisible to TeX
\providecommand\bibfield[2]{#2}
\providecommand\bibinfo[2]{#2}
\providecommand\natexlab[1]{#1}
\providecommand\showeprint[2][]{arXiv:#2}

\bibitem[\protect\citeauthoryear{??}{Fol}{2009}]%
        {FollowSWM09}
 \bibinfo{year}{2009}\natexlab{}.
\newblock \bibinfo{title}{Follow the Sun, Wind, Moon}.
\newblock
\newblock
\newblock
\shownote{\url{https://www.vertatique.com/cloud-computing-starting-follow-sunwindmoon}}.


\bibitem[\protect\citeauthoryear{??}{Lan}{2018}]%
        {Lancium}
 \bibinfo{year}{2018}\natexlab{}.
\newblock \bibinfo{title}{Lancium}.
\newblock \bibinfo{howpublished}{\url{https://www.lancium.com}}.
\newblock
\newblock
\shownote{A startup company, building zero-carbon cloud computing resources.}.


\bibitem[\protect\citeauthoryear{??}{Gre}{2019}]%
        {Greenpeace-Nova19}
 \bibinfo{year}{2019}\natexlab{}.
\newblock \bibinfo{title}{{Clicking Clean Virginia: The Dirty Energy Powering
  Data Center Alley}}.
\newblock
\newblock
\newblock
\shownote{\url{https://www.greenpeace.org/usa/reports/click-clean-virginia/}}.


\bibitem[\protect\citeauthoryear{Acun, Lee, Maeng, Chakkaravarthy, Gupta,
  Brooks, and Wu}{Acun et~al\mbox{.}}{2022}]%
        {acun2022holistic}
\bibfield{author}{\bibinfo{person}{Bilge Acun}, \bibinfo{person}{Benjamin Lee},
  \bibinfo{person}{Kiwan Maeng}, \bibinfo{person}{Manoj Chakkaravarthy},
  \bibinfo{person}{Udit Gupta}, \bibinfo{person}{David Brooks}, {and}
  \bibinfo{person}{Carole-Jean Wu}.} \bibinfo{year}{2022}\natexlab{}.
\newblock \showarticletitle{A Holistic Approach for Designing Carbon Aware
  Datacenters}.
\newblock \bibinfo{journal}{\emph{arXiv preprint arXiv:2201.10036}}
  (\bibinfo{year}{2022}).
\newblock


\bibitem[\protect\citeauthoryear{Administration}{Administration}{2022}]%
        {eiaPowerPrice}
\bibfield{author}{\bibinfo{person}{U.S. Energy~Information Administration}.}
  \bibinfo{year}{2022}\natexlab{}.
\newblock \bibinfo{title}{Electric Power Monthly}.
\newblock
\newblock
\urldef\tempurl%
\url{https://www.eia.gov/electricity/monthly/epm_table_grapher.php?t=epmt_5_03}
\showURL{%
\tempurl}


\bibitem[\protect\citeauthoryear{Agency}{Agency}{2020}]%
        {eGrid}
\bibfield{author}{\bibinfo{person}{United States Environmental~Protection
  Agency}.} \bibinfo{year}{2020}\natexlab{}.
\newblock \bibinfo{title}{Emissions \& Generation Resource Integrated
  Database}.
\newblock
\newblock
\urldef\tempurl%
\url{https://www.epa.gov/egrid/data-explorer}
\showURL{%
\tempurl}


\bibitem[\protect\citeauthoryear{Ahmad, Rosenthal, Hajiesmaili, and
  Sitaraman}{Ahmad et~al\mbox{.}}{2019}]%
        {Rosenthal19}
\bibfield{author}{\bibinfo{person}{Sohaib Ahmad}, \bibinfo{person}{Arielle
  Rosenthal}, \bibinfo{person}{Mohammad~H. Hajiesmaili}, {and}
  \bibinfo{person}{Ramesh~K. Sitaraman}.} \bibinfo{year}{2019}\natexlab{}.
\newblock \showarticletitle{Learning from Optimal: Energy Procurement
  Strategies for Data Centers}. In \bibinfo{booktitle}{\emph{Proceedings of the
  Tenth ACM International Conference on Future Energy Systems}} (Phoenix, AZ,
  USA) \emph{(\bibinfo{series}{e-Energy '19})}. \bibinfo{publisher}{Association
  for Computing Machinery}, \bibinfo{address}{New York, NY, USA},
  \bibinfo{pages}{326–330}.
\newblock
\showISBNx{9781450366717}
\urldef\tempurl%
\url{https://doi.org/10.1145/3307772.3328308}
\showDOI{\tempurl}


\bibitem[\protect\citeauthoryear{Amazon}{Amazon}{2022}]%
        {amazonEnv}
\bibfield{author}{\bibinfo{person}{Amazon}.} \bibinfo{year}{2022}\natexlab{}.
\newblock \bibinfo{title}{2021 Environmental Report}.
\newblock
\newblock
\newblock
\shownote{Available from
  \url{https://sustainability.aboutamazon.com/reporting-and-downloads}}.


\bibitem[\protect\citeauthoryear{(AWS)}{(AWS)}{[n.\,d.]}]%
        {AmazonCloudDatacenters}
\bibfield{author}{\bibinfo{person}{Amazon Web~Services (AWS)}.}
  \bibinfo{year}{[n.\,d.]}\natexlab{}.
\newblock \bibinfo{title}{Amazon Global Datacenters}.
\newblock
\newblock


\bibitem[\protect\citeauthoryear{Bashir, Guo, Hajiesmaili, Irwin, Shenoy,
  Sitaraman, Souza, and Wierman}{Bashir et~al\mbox{.}}{2021}]%
        {bashir2021enabling}
\bibfield{author}{\bibinfo{person}{Noman Bashir}, \bibinfo{person}{Tian Guo},
  \bibinfo{person}{Mohammad Hajiesmaili}, \bibinfo{person}{David Irwin},
  \bibinfo{person}{Prashant Shenoy}, \bibinfo{person}{Ramesh Sitaraman},
  \bibinfo{person}{Abel Souza}, {and} \bibinfo{person}{Adam Wierman}.}
  \bibinfo{year}{2021}\natexlab{}.
\newblock \showarticletitle{Enabling Sustainable Clouds: The Case for
  Virtualizing the Energy System}. In \bibinfo{booktitle}{\emph{Proceedings of
  the ACM Symposium on Cloud Computing}}. \bibinfo{pages}{350--358}.
\newblock


\bibitem[\protect\citeauthoryear{Bird, Milligan, and Lew}{Bird
  et~al\mbox{.}}{2013}]%
        {Bird2013}
\bibfield{author}{\bibinfo{person}{Lori Bird}, \bibinfo{person}{M Milligan},
  {and} \bibinfo{person}{Debra Lew}.} \bibinfo{year}{2013}\natexlab{}.
\newblock \bibinfo{booktitle}{\emph{{Integrating Variable Renewable Energy:
  Challenges and Solutions}}}.
\newblock \bibinfo{type}{{T}echnical {R}eport}. \bibinfo{institution}{NREL}.
\newblock


\bibitem[\protect\citeauthoryear{Bloomberg}{Bloomberg}{2018}]%
        {EU-RPS45}
\bibfield{author}{\bibinfo{person}{Bloomberg}.}
  \bibinfo{year}{2018}\natexlab{}.
\newblock \showarticletitle{European Union Aims to Be First Carbon Neutral
  Major Economy by 2050}.
\newblock \bibinfo{journal}{\emph{Fortune}} (\bibinfo{date}{November}
  \bibinfo{year}{2018}).
\newblock


\bibitem[\protect\citeauthoryear{Campbell}{Campbell}{2022}]%
        {irelandDC14percent}
\bibfield{author}{\bibinfo{person}{John Campbell}.}
  \bibinfo{year}{2022}\natexlab{}.
\newblock \bibinfo{title}{Data centres used 14\% of Republic of Ireland's
  electricity use}.
\newblock
\newblock
\urldef\tempurl%
\url{https://www.bbc.com/news/world-europe-61308747}
\showURL{%
\tempurl}


\bibitem[\protect\citeauthoryear{Chen, Ren, Ren, and Wierman}{Chen
  et~al\mbox{.}}{2015}]%
        {chen2015greening}
\bibfield{author}{\bibinfo{person}{Niangjun Chen}, \bibinfo{person}{Xiaoqi
  Ren}, \bibinfo{person}{Shaolei Ren}, {and} \bibinfo{person}{Adam Wierman}.}
  \bibinfo{year}{2015}\natexlab{}.
\newblock \showarticletitle{Greening multi-tenant data center demand response}.
\newblock \bibinfo{journal}{\emph{Performance Evaluation}}
  \bibinfo{volume}{91} (\bibinfo{year}{2015}), \bibinfo{pages}{229--254}.
\newblock


\bibitem[\protect\citeauthoryear{Chen, Jiao, Wang, and Liu}{Chen
  et~al\mbox{.}}{2019}]%
        {chen2019online}
\bibfield{author}{\bibinfo{person}{Shutong Chen}, \bibinfo{person}{Lei Jiao},
  \bibinfo{person}{Lin Wang}, {and} \bibinfo{person}{Fangming Liu}.}
  \bibinfo{year}{2019}\natexlab{}.
\newblock \showarticletitle{An online market mechanism for edge emergency
  demand response via cloudlet control}. In \bibinfo{booktitle}{\emph{IEEE
  INFOCOM 2019-IEEE Conference on Computer Communications}}. IEEE,
  \bibinfo{pages}{2566--2574}.
\newblock


\bibitem[\protect\citeauthoryear{Chien}{Chien}{2020}]%
        {CAISO-stranded18}
\bibfield{author}{\bibinfo{person}{Andrew~A Chien}.}
  \bibinfo{year}{2020}\natexlab{}.
\newblock \showarticletitle{Characterizing Opportunity Power in the California
  Independent System Operator (CAISO) in Years 2015-2017}.
\newblock \bibinfo{journal}{\emph{Energy and Earth Science}}
  \bibinfo{volume}{3}, \bibinfo{number}{2} (\bibinfo{date}{December}
  \bibinfo{year}{2020}).
\newblock
\newblock
\shownote{Also available as University of Chicago, Computer Science TR-2018-07,
  \url{https://newtraell.cs.uchicago.edu/research/publications/techreports}}.


\bibitem[\protect\citeauthoryear{Chien, Wolski, and Yang}{Chien
  et~al\mbox{.}}{2015}]%
        {chien2015zero}
\bibfield{author}{\bibinfo{person}{Andrew~A Chien}, \bibinfo{person}{Richard
  Wolski}, {and} \bibinfo{person}{Fan Yang}.} \bibinfo{year}{2015}\natexlab{}.
\newblock \showarticletitle{The Zero-Carbon Cloud: High-Value, Dispatchable
  Demand for Renewable Power Generators}.
\newblock \bibinfo{journal}{\emph{The Electricity Journal}}
  (\bibinfo{year}{2015}), \bibinfo{pages}{110--118}.
\newblock


\bibitem[\protect\citeauthoryear{Chien, Yang, and Zhang}{Chien
  et~al\mbox{.}}{2018}]%
        {AIMS18}
\bibfield{author}{\bibinfo{person}{Andrew~A. Chien}, \bibinfo{person}{Fan
  Yang}, {and} \bibinfo{person}{Chaojie Zhang}.}
  \bibinfo{year}{2018}\natexlab{}.
\newblock \showarticletitle{Characterizing Curtailed and Uneconomic Renewable
  Power in the Mid-continent Independent System Operator}.
\newblock \bibinfo{journal}{\emph{AIMS Energy}} \bibinfo{volume}{6},
  \bibinfo{number}{2} (\bibinfo{date}{December} \bibinfo{year}{2018}),
  \bibinfo{pages}{376--401}.
\newblock


\bibitem[\protect\citeauthoryear{Corporation}{Corporation}{2021}]%
        {microsoftEnv}
\bibfield{author}{\bibinfo{person}{Microsoft Corporation}.}
  \bibinfo{year}{2021}\natexlab{}.
\newblock \bibinfo{title}{Microsft 2021 Environmental Sustainability Report}.
\newblock
\newblock
\newblock
\shownote{Available from
  \url{https://www.microsoft.com/en-us/corporate-responsibility/sustainability/report}}.


\bibitem[\protect\citeauthoryear{Corradi}{Corradi}{2022}]%
        {tomorrowCarbonIntensityBlog}
\bibfield{author}{\bibinfo{person}{Olivier Corradi}.}
  \bibinfo{year}{2022}\natexlab{}.
\newblock \bibinfo{title}{Marginal vs average: which one to use in practice?}
\newblock
\newblock
\urldef\tempurl%
\url{https://www.electricitymaps.com/blog/marginal-vs-average-real-time-decision-making}
\showURL{%
\tempurl}


\bibitem[\protect\citeauthoryear{Cortez, Bonde, Muzio, Russinovich, Fontoura,
  and Bianchini}{Cortez et~al\mbox{.}}{2017}]%
        {ResourceCentral17}
\bibfield{author}{\bibinfo{person}{Eli Cortez}, \bibinfo{person}{Anand Bonde},
  \bibinfo{person}{Alexandre Muzio}, \bibinfo{person}{Mark Russinovich},
  \bibinfo{person}{Marcus Fontoura}, {and} \bibinfo{person}{Ricardo
  Bianchini}.} \bibinfo{year}{2017}\natexlab{}.
\newblock \showarticletitle{Resource Central: Understanding and Predicting
  Workloads for Improved Resource Management in Large Cloud Platforms}. In
  \bibinfo{booktitle}{\emph{Proceedings of the 26th Symposium on Operating
  Systems Principles}} \emph{(\bibinfo{series}{SOSP '17})}.
  \bibinfo{pages}{153--167}.
\newblock
\urldef\tempurl%
\url{https://doi.org/10.1145/3132747.3132772}
\showDOI{\tempurl}


\bibitem[\protect\citeauthoryear{CPUC}{CPUC}{[n.\,d.]}]%
        {rps-california}
\bibfield{author}{\bibinfo{person}{CPUC}.} \bibinfo{year}{[n.\,d.]}\natexlab{}.
\newblock \bibinfo{title}{{California Public Utilities Commission (CPUC)}}.
\newblock
\newblock
\newblock
\shownote{\url{https://www.cpuc.ca.gov/rps}}.


\bibitem[\protect\citeauthoryear{Deng, Liu, Jin, Li, and Li}{Deng
  et~al\mbox{.}}{2014}]%
        {deng2014harnessing}
\bibfield{author}{\bibinfo{person}{Wei Deng}, \bibinfo{person}{Fangming Liu},
  \bibinfo{person}{Hai Jin}, \bibinfo{person}{Bo Li}, {and}
  \bibinfo{person}{Dan Li}.} \bibinfo{year}{2014}\natexlab{}.
\newblock \showarticletitle{Harnessing renewable energy in cloud datacenters:
  opportunities and challenges}.
\newblock \bibinfo{journal}{\emph{iEEE Network}} \bibinfo{volume}{28},
  \bibinfo{number}{1} (\bibinfo{year}{2014}), \bibinfo{pages}{48--55}.
\newblock


\bibitem[\protect\citeauthoryear{Dou, Qi, Wei, and Song}{Dou
  et~al\mbox{.}}{2017}]%
        {dou2017carbon}
\bibfield{author}{\bibinfo{person}{Hui Dou}, \bibinfo{person}{Yong Qi},
  \bibinfo{person}{Wei Wei}, {and} \bibinfo{person}{Houbing Song}.}
  \bibinfo{year}{2017}\natexlab{}.
\newblock \showarticletitle{Carbon-aware electricity cost minimization for
  sustainable data centers}.
\newblock \bibinfo{journal}{\emph{IEEE Transactions on Sustainable Computing}}
  \bibinfo{volume}{2}, \bibinfo{number}{2} (\bibinfo{year}{2017}),
  \bibinfo{pages}{211--223}.
\newblock


\bibitem[\protect\citeauthoryear{Driscoll}{Driscoll}{2022}]%
        {California-rps-90}
\bibfield{author}{\bibinfo{person}{William Driscoll}.}
  \bibinfo{year}{2022}\natexlab{}.
\newblock \bibinfo{title}{California law would target 90\% renewable and
  zero-carbon electricity by 2035}.
\newblock
\newblock
\urldef\tempurl%
\url{https://pv-magazine-usa.com/2022/09/06/california-law-would-target-90-renewable-and-zero-carbon-electricity-by-2035/}
\showURL{%
\tempurl}


\bibitem[\protect\citeauthoryear{Dunning, Huchette, and Lubin}{Dunning
  et~al\mbox{.}}{2017}]%
        {dunning2017jump}
\bibfield{author}{\bibinfo{person}{Iain Dunning}, \bibinfo{person}{Joey
  Huchette}, {and} \bibinfo{person}{Miles Lubin}.}
  \bibinfo{year}{2017}\natexlab{}.
\newblock \showarticletitle{JuMP: A modeling language for mathematical
  optimization}.
\newblock \bibinfo{journal}{\emph{SIAM review}} \bibinfo{volume}{59},
  \bibinfo{number}{2} (\bibinfo{year}{2017}), \bibinfo{pages}{295--320}.
\newblock


\bibitem[\protect\citeauthoryear{Energy}{Energy}{2018}]%
        {Dominion18IRP}
\bibfield{author}{\bibinfo{person}{Dominion Energy}.}
  \bibinfo{year}{2018}\natexlab{}.
\newblock \bibinfo{title}{2018 Virginia Integrated Resource Plan}.
\newblock
\newblock
\urldef\tempurl%
\url{https://rga.lis.virginia.gov/Published/2018/RD249}
\showURL{%
\tempurl}


\bibitem[\protect\citeauthoryear{Energy}{Energy}{2020}]%
        {dominion20IRP}
\bibfield{author}{\bibinfo{person}{Dominion Energy}.}
  \bibinfo{year}{2020}\natexlab{}.
\newblock \bibinfo{title}{2020 Virginia Integrated Resource Plan}.
\newblock
\newblock
\urldef\tempurl%
\url{https://www.dominionenergy.com/-/media/pdfs/global/2020-va-integrated-resource-plan.pdf}
\showURL{%
\tempurl}


\bibitem[\protect\citeauthoryear{Energy}{Energy}{2021}]%
        {dominion21IRP}
\bibfield{author}{\bibinfo{person}{Dominion Energy}.}
  \bibinfo{year}{2021}\natexlab{}.
\newblock \bibinfo{title}{2021 Update to the 2020 Integrated Resource Plan}.
\newblock
\newblock
\urldef\tempurl%
\url{https://www.dominionenergy.com/-/media/pdfs/global/company/2021-de-integrated-resource-plan.pdf}
\showURL{%
\tempurl}


\bibitem[\protect\citeauthoryear{Foley}{Foley}{2022}]%
        {Microsoft-30pct}
\bibfield{author}{\bibinfo{person}{Mary~Jo Foley}.}
  \bibinfo{year}{2022}\natexlab{}.
\newblock \bibinfo{title}{Cloud revenues power Microsoft's \$51.7 billion Q2 in
  fiscal year 2022}.
\newblock
\newblock
\newblock
\shownote{\url{https://www.zdnet.com/article/microsoft-cloud-revenues-power-microsofts-51-7-billion-second-fy22-quarter/}}.


\bibitem[\protect\citeauthoryear{Galvin}{Galvin}{2021}]%
        {irelandDC}
\bibfield{author}{\bibinfo{person}{Robbie Galvin}.}
  \bibinfo{year}{2021}\natexlab{}.
\newblock \bibinfo{title}{Data Centers Are Pushing Ireland’s Electric Grid to
  the Brink}.
\newblock
\newblock
\urldef\tempurl%
\url{https://gizmodo.com/data-centers-are-pushing-ireland-s-electric-grid-to-the-1848282390}
\showURL{%
\tempurl}


\bibitem[\protect\citeauthoryear{Goiri, Katsak, Le, Nguyen, and
  Bianchini}{Goiri et~al\mbox{.}}{2013}]%
        {goiri2013parasol}
\bibfield{author}{\bibinfo{person}{{\'I}{\~n}igo Goiri},
  \bibinfo{person}{William Katsak}, \bibinfo{person}{Kien Le},
  \bibinfo{person}{Thu~D Nguyen}, {and} \bibinfo{person}{Ricardo Bianchini}.}
  \bibinfo{year}{2013}\natexlab{}.
\newblock \showarticletitle{{Parasol and greenswitch: Managing datacenters
  powered by renewable energy}}. In \bibinfo{booktitle}{\emph{ACM SIGARCH
  Computer Architecture News}}. ACM, \bibinfo{pages}{51--64}.
\newblock


\bibitem[\protect\citeauthoryear{Golin and Pearson}{Golin and Pearson}{2022}]%
        {googlePolicyRoadmap}
\bibfield{author}{\bibinfo{person}{Caroline Golin} {and} \bibinfo{person}{Nick
  Pearson}.} \bibinfo{year}{2022}\natexlab{}.
\newblock \bibinfo{title}{A policy roadmap for 24/7 carbon-free energy}.
\newblock
\newblock
\urldef\tempurl%
\url{https://cloud.google.com/blog/topics/sustainability/a-policy-roadmap-for-achieving-247-carbon-free-energy}
\showURL{%
\tempurl}


\bibitem[\protect\citeauthoryear{{Google}}{{Google}}{[n.\,d.]}]%
        {GoogleCloudDatacenters}
\bibfield{author}{\bibinfo{person}{{Google}}.}
  \bibinfo{year}{[n.\,d.]}\natexlab{}.
\newblock \bibinfo{title}{About Google Datacenters}.
\newblock
\newblock
\newblock
\shownote{\url{https://www.google.com/about/datacenters/}}.


\bibitem[\protect\citeauthoryear{Google}{Google}{2018}]%
        {GoogleWhite18}
\bibfield{author}{\bibinfo{person}{Google}.} \bibinfo{year}{2018}\natexlab{}.
\newblock \bibinfo{booktitle}{\emph{Moving toward 24x7 Carbon-Free Energy at
  Google Data Centers: Progress and Insights}}.
\newblock \bibinfo{type}{{T}echnical {R}eport}. \bibinfo{institution}{Google}.
\newblock


\bibitem[\protect\citeauthoryear{Google}{Google}{2022}]%
        {googleEnv}
\bibfield{author}{\bibinfo{person}{Google}.} \bibinfo{year}{2022}\natexlab{}.
\newblock \bibinfo{title}{Google 2021 Environmental Report}.
\newblock
\newblock
\newblock
\shownote{Available from
  \url{https://www.gstatic.com/gumdrop/sustainability/google-2022-environmental-report.pdf}}.


\bibitem[\protect\citeauthoryear{{Gupta}, {Shenoy}, and {Sitaraman}}{{Gupta}
  et~al\mbox{.}}{2018}]%
        {Sitaraman-Solar18}
\bibfield{author}{\bibinfo{person}{V. {Gupta}}, \bibinfo{person}{P. {Shenoy}},
  {and} \bibinfo{person}{R.~K. {Sitaraman}}.} \bibinfo{year}{2018}\natexlab{}.
\newblock \showarticletitle{Efficient solar provisioning for net-zero
  Internet-scale distributed networks}. In \bibinfo{booktitle}{\emph{2018 10th
  International Conference on Communication Systems Networks (COMSNETS)}}.
  \bibinfo{pages}{372--379}.
\newblock


\bibitem[\protect\citeauthoryear{Gurobi~Optimization}{Gurobi~Optimization}{2020}]%
        {gurobi}
\bibfield{author}{\bibinfo{person}{LLC Gurobi~Optimization}.}
  \bibinfo{year}{2020}\natexlab{}.
\newblock \bibinfo{title}{Gurobi Optimizer Reference Manual}.
\newblock
\newblock
\urldef\tempurl%
\url{http://www.gurobi.com}
\showURL{%
\tempurl}


\bibitem[\protect\citeauthoryear{GWEC}{GWEC}{2016}]%
        {GWEC-Annual16}
\bibfield{author}{\bibinfo{person}{GWEC}.} \bibinfo{year}{2016}\natexlab{}.
\newblock \bibinfo{booktitle}{\emph{Global Wind Report: Annual Market Update}}.
\newblock \bibinfo{type}{{T}echnical {R}eport}. \bibinfo{institution}{Global
  Wind Energy Council}.
\newblock
\newblock
\shownote{Documents curtailment around the world}.


\bibitem[\protect\citeauthoryear{Han}{Han}{2015}]%
        {ChinaWilson15}
\bibfield{author}{\bibinfo{person}{Siqi Han}.} \bibinfo{year}{2015}\natexlab{}.
\newblock \bibinfo{title}{The Wind is Wasted in China}.
\newblock \bibinfo{howpublished}{\url{https://www.wilsoncenter.org/}}.
\newblock


\bibitem[\protect\citeauthoryear{Imboden}{Imboden}{2021}]%
        {dcMarketReport}
\bibfield{author}{\bibinfo{person}{Kevin Imboden}.}
  \bibinfo{year}{2021}\natexlab{}.
\newblock \bibinfo{title}{2022 Global Data Center Market Comparison}.
\newblock
\newblock
\urldef\tempurl%
\url{https://cushwake.cld.bz/2022-Global-Data-Center-Market-Comparison}
\showURL{%
\tempurl}


\bibitem[\protect\citeauthoryear{ISO}{ISO}{2016}]%
        {caisoGHG}
\bibfield{author}{\bibinfo{person}{California ISO}.}
  \bibinfo{year}{2016}\natexlab{}.
\newblock \bibinfo{title}{Greenhouse Gas Emission Tracking Methodology}.
\newblock
\newblock
\urldef\tempurl%
\url{https://www.caiso.com/Documents/GreenhouseGasEmissionsTracking-Methodology.pdf}
\showURL{%
\tempurl}


\bibitem[\protect\citeauthoryear{ISO}{ISO}{2022a}]%
        {caisoManagingOversupply}
\bibfield{author}{\bibinfo{person}{California ISO}.}
  \bibinfo{year}{2022}\natexlab{a}.
\newblock \bibinfo{title}{Managing Oversupply}.
\newblock
\newblock
\urldef\tempurl%
\url{https://www.caiso.com/informed/Pages/ManagingOversupply.aspx}
\showURL{%
\tempurl}


\bibitem[\protect\citeauthoryear{ISO}{ISO}{2022b}]%
        {caisoMarketProcess}
\bibfield{author}{\bibinfo{person}{California ISO}.}
  \bibinfo{year}{2022}\natexlab{b}.
\newblock \bibinfo{title}{Market Processes and Products}.
\newblock
\newblock
\urldef\tempurl%
\url{http://www.caiso.com/market/Pages/MarketProcesses.aspx}
\showURL{%
\tempurl}


\bibitem[\protect\citeauthoryear{Jones}{Jones}{2018}]%
        {Nature18}
\bibfield{author}{\bibinfo{person}{Nicola Jones}.}
  \bibinfo{year}{2018}\natexlab{}.
\newblock \showarticletitle{How to Stop Data Centres from Gobbling up the
  World's Electricity}.
\newblock \bibinfo{journal}{\emph{Nature}} (\bibinfo{date}{September}
  \bibinfo{year}{2018}).
\newblock


\bibitem[\protect\citeauthoryear{Joppa}{Joppa}{2021}]%
        {msftSustainability2021}
\bibfield{author}{\bibinfo{person}{Lucas Joppa}.}
  \bibinfo{year}{2021}\natexlab{}.
\newblock \bibinfo{title}{Made to measure: Sustainability commitment progress
  and updates}.
\newblock
\newblock
\urldef\tempurl%
\url{https://blogs.microsoft.com/blog/2021/07/14/made-to-measure-sustainability-commitment-progress-and-updates/}
\showURL{%
\tempurl}


\bibitem[\protect\citeauthoryear{Judge}{Judge}{2022a}]%
        {NoVAHaltDC}
\bibfield{author}{\bibinfo{person}{Peter Judge}.}
  \bibinfo{year}{2022}\natexlab{a}.
\newblock \bibinfo{title}{Dominion Energy admits it can't meet data center
  power demands in Virginia}.
\newblock
\newblock
\urldef\tempurl%
\url{https://www.datacenterdynamics.com/en/news/dominion-energy-admits-it-cant-meet-data-center-power-demands-in-virginia/}
\showURL{%
\tempurl}


\bibitem[\protect\citeauthoryear{Judge}{Judge}{2022b}]%
        {IrelandHaltDC}
\bibfield{author}{\bibinfo{person}{Peter Judge}.}
  \bibinfo{year}{2022}\natexlab{b}.
\newblock \bibinfo{title}{EirGrid pulls plug on 30 Irish data center projects}.
\newblock
\newblock
\urldef\tempurl%
\url{https://www.datacenterdynamics.com/en/news/eirgrid-pulls-plug-on-30-irish-data-center-projects/}
\showURL{%
\tempurl}


\bibitem[\protect\citeauthoryear{Judge}{Judge}{2022c}]%
        {longLastingBattery}
\bibfield{author}{\bibinfo{person}{Peter Judge}.}
  \bibinfo{year}{2022}\natexlab{c}.
\newblock \bibinfo{title}{Google and Microsoft join long term energy storage
  group}.
\newblock
\newblock
\urldef\tempurl%
\url{https://www.datacenterdynamics.com/en/news/google-and-microsoft-join-long-term-energy-storage-group/}
\showURL{%
\tempurl}


\bibitem[\protect\citeauthoryear{Kim, Yang, Zavala, and Chien}{Kim
  et~al\mbox{.}}{2016}]%
        {KYZC2016}
\bibfield{author}{\bibinfo{person}{Kibaek Kim}, \bibinfo{person}{Fan Yang},
  \bibinfo{person}{Victor Zavala}, {and} \bibinfo{person}{Andrew~A. Chien}.}
  \bibinfo{year}{2016}\natexlab{}.
\newblock \showarticletitle{Data Centers as Dispatchable Loads to Harness
  Stranded Power}.
\newblock \bibinfo{journal}{\emph{IEEE Transactions on Sustainable Energy}}
  (\bibinfo{year}{2016}).
\newblock
\newblock
\shownote{DOI 10.1109/TSTE.2016.2593607}.


\bibitem[\protect\citeauthoryear{Kwon and Han}{Kwon and Han}{2006}]%
        {kwon2006receding}
\bibfield{author}{\bibinfo{person}{Wook~Hyun Kwon} {and}
  \bibinfo{person}{Soo~Hee Han}.} \bibinfo{year}{2006}\natexlab{}.
\newblock \bibinfo{booktitle}{\emph{Receding horizon control: model predictive
  control for state models}}.
\newblock \bibinfo{publisher}{Springer Science \& Business Media}.
\newblock


\bibitem[\protect\citeauthoryear{Le, Bianchini, Nguyen, Bilgir, and
  Martonosi}{Le et~al\mbox{.}}{2010}]%
        {Kien10}
\bibfield{author}{\bibinfo{person}{Kien Le}, \bibinfo{person}{Ricardo
  Bianchini}, \bibinfo{person}{Thu~D. Nguyen}, \bibinfo{person}{Ozlem Bilgir},
  {and} \bibinfo{person}{Margaret Martonosi}.} \bibinfo{year}{2010}\natexlab{}.
\newblock \showarticletitle{Capping the Brown Energy Consumption of Internet
  Services at Low Cost}. In \bibinfo{booktitle}{\emph{Proceedings of the
  International Conference on Green Computing}}
  \emph{(\bibinfo{series}{GREENCOMP '10})}. \bibinfo{publisher}{IEEE Computer
  Society}, \bibinfo{address}{USA}, \bibinfo{pages}{3–14}.
\newblock
\showISBNx{9781424476121}
\urldef\tempurl%
\url{https://doi.org/10.1109/GREENCOMP.2010.5598305}
\showDOI{\tempurl}


\bibitem[\protect\citeauthoryear{Le, Liu, Chen, and Bash}{Le
  et~al\mbox{.}}{2016}]%
        {le2016joint}
\bibfield{author}{\bibinfo{person}{Tan~N Le}, \bibinfo{person}{Zhenhua Liu},
  \bibinfo{person}{Yuan Chen}, {and} \bibinfo{person}{Cullen Bash}.}
  \bibinfo{year}{2016}\natexlab{}.
\newblock \showarticletitle{Joint capacity planning and operational management
  for sustainable data centers and demand response}. In
  \bibinfo{booktitle}{\emph{Proceedings of the Seventh International Conference
  on Future Energy Systems}}. \bibinfo{pages}{1--12}.
\newblock


\bibitem[\protect\citeauthoryear{Lewis}{Lewis}{2023}]%
        {ppa2022}
\bibfield{author}{\bibinfo{person}{Michelle Lewis}.}
  \bibinfo{year}{2023}\natexlab{}.
\newblock \bibinfo{title}{These big tech firms bought the most clean energy in
  2022}.
\newblock
\newblock
\newblock
\shownote{Available from
  \url{https://electrek.co/2023/02/09/big-tech-clean-energy-2022/}}.


\bibitem[\protect\citeauthoryear{Lin and Chien}{Lin and Chien}{2020}]%
        {ERCOT-stranded20}
\bibfield{author}{\bibinfo{person}{Liuzixuan Lin} {and}
  \bibinfo{person}{Andrew~A. Chien}.} \bibinfo{year}{2020}\natexlab{}.
\newblock \bibinfo{booktitle}{\emph{Characterizing Stranded Power in the ERCOT
  in Years 2012-2019: A Preliminary Report}}.
\newblock \bibinfo{type}{{T}echnical {R}eport} TR-2020-06.
  \bibinfo{institution}{University of Chicago}.
\newblock


\bibitem[\protect\citeauthoryear{Lin, Zavala, and Chien}{Lin
  et~al\mbox{.}}{2021}]%
        {lin2021evaluating}
\bibfield{author}{\bibinfo{person}{Liuzixuan Lin}, \bibinfo{person}{Victor~M
  Zavala}, {and} \bibinfo{person}{Andrew~A Chien}.}
  \bibinfo{year}{2021}\natexlab{}.
\newblock \showarticletitle{Evaluating Coupling Models for Cloud Datacenters
  and Power Grids}. In \bibinfo{booktitle}{\emph{Proceedings of the Twelfth ACM
  International Conference on Future Energy Systems}}.
  \bibinfo{pages}{171--184}.
\newblock


\bibitem[\protect\citeauthoryear{Lin, Wierman, Andrew, and Thereska}{Lin
  et~al\mbox{.}}{2012}]%
        {lin2012dynamic}
\bibfield{author}{\bibinfo{person}{Minghong Lin}, \bibinfo{person}{Adam
  Wierman}, \bibinfo{person}{Lachlan~LH Andrew}, {and} \bibinfo{person}{Eno
  Thereska}.} \bibinfo{year}{2012}\natexlab{}.
\newblock \showarticletitle{Dynamic right-sizing for power-proportional data
  centers}.
\newblock \bibinfo{journal}{\emph{IEEE/ACM Transactions on Networking}}
  \bibinfo{volume}{21}, \bibinfo{number}{5} (\bibinfo{year}{2012}),
  \bibinfo{pages}{1378--1391}.
\newblock


\bibitem[\protect\citeauthoryear{Lindberg, Abdennadher, Chen, Lesieutre, and
  Roald}{Lindberg et~al\mbox{.}}{2021}]%
        {lindberg2021guide}
\bibfield{author}{\bibinfo{person}{Julia Lindberg}, \bibinfo{person}{Yasmine
  Abdennadher}, \bibinfo{person}{Jiaqi Chen}, \bibinfo{person}{Bernard~C
  Lesieutre}, {and} \bibinfo{person}{Line Roald}.}
  \bibinfo{year}{2021}\natexlab{}.
\newblock \showarticletitle{A Guide to Reducing Carbon Emissions through Data
  Center Geographical Load Shifting}. In \bibinfo{booktitle}{\emph{Proceedings
  of the Twelfth ACM International Conference on Future Energy Systems}}.
  \bibinfo{pages}{430--436}.
\newblock


\bibitem[\protect\citeauthoryear{Lindberg, Roald, and Lesieutre}{Lindberg
  et~al\mbox{.}}{2020}]%
        {lindberg2020environmental}
\bibfield{author}{\bibinfo{person}{Julia Lindberg}, \bibinfo{person}{Line
  Roald}, {and} \bibinfo{person}{Bernard Lesieutre}.}
  \bibinfo{year}{2020}\natexlab{}.
\newblock \showarticletitle{The Environmental Potential of Hyper-Scale Data
  Centers: Using Locational Marginal CO2 Emissions to Guide Geographical Load
  Shifting}. In \bibinfo{booktitle}{\emph{Proceedings of the 54th Hawaii
  International Conference on System Sciences}}. \bibinfo{pages}{3158}.
\newblock


\bibitem[\protect\citeauthoryear{Liu, Chen, Bash, Wierman, Gmach, Wang, Marwah,
  and Hyser}{Liu et~al\mbox{.}}{2012}]%
        {liu2012renewable}
\bibfield{author}{\bibinfo{person}{Zhenhua Liu}, \bibinfo{person}{Yuan Chen},
  \bibinfo{person}{Cullen Bash}, \bibinfo{person}{Adam Wierman},
  \bibinfo{person}{Daniel Gmach}, \bibinfo{person}{Zhikui Wang},
  \bibinfo{person}{Manish Marwah}, {and} \bibinfo{person}{Chris Hyser}.}
  \bibinfo{year}{2012}\natexlab{}.
\newblock \showarticletitle{Renewable and cooling aware workload management for
  sustainable data centers}. In \bibinfo{booktitle}{\emph{Proceedings of the
  12th ACM SIGMETRICS/PERFORMANCE joint international conference on Measurement
  and Modeling of Computer Systems}}. \bibinfo{pages}{175--186}.
\newblock


\bibitem[\protect\citeauthoryear{{Liu}, {Lin}, {Wierman}, {Low}, and
  {Andrew}}{{Liu} et~al\mbox{.}}{2015}]%
        {Liu-Greening15}
\bibfield{author}{\bibinfo{person}{Z. {Liu}}, \bibinfo{person}{M. {Lin}},
  \bibinfo{person}{A. {Wierman}}, \bibinfo{person}{S. {Low}}, {and}
  \bibinfo{person}{L.~L.~H. {Andrew}}.} \bibinfo{year}{2015}\natexlab{}.
\newblock \showarticletitle{Greening Geographical Load Balancing}.
\newblock \bibinfo{journal}{\emph{IEEE/ACM Transactions on Networking}}
  \bibinfo{volume}{23}, \bibinfo{number}{2} (\bibinfo{year}{2015}),
  \bibinfo{pages}{657--671}.
\newblock


\bibitem[\protect\citeauthoryear{Liu, Wierman, Chen, Razon, and Chen}{Liu
  et~al\mbox{.}}{2013}]%
        {liu2013data}
\bibfield{author}{\bibinfo{person}{Zhenhua Liu}, \bibinfo{person}{Adam
  Wierman}, \bibinfo{person}{Yuan Chen}, \bibinfo{person}{Benjamin Razon},
  {and} \bibinfo{person}{Niangjun Chen}.} \bibinfo{year}{2013}\natexlab{}.
\newblock \showarticletitle{Data center demand response: Avoiding the
  coincident peak via workload shifting and local generation}.
\newblock \bibinfo{journal}{\emph{Performance Evaluation}}
  \bibinfo{volume}{70}, \bibinfo{number}{10} (\bibinfo{year}{2013}),
  \bibinfo{pages}{770--791}.
\newblock


\bibitem[\protect\citeauthoryear{Luo, Rao, and Liu}{Luo et~al\mbox{.}}{2013}]%
        {luo2013temporal}
\bibfield{author}{\bibinfo{person}{Jianying Luo}, \bibinfo{person}{Lei Rao},
  {and} \bibinfo{person}{Xue Liu}.} \bibinfo{year}{2013}\natexlab{}.
\newblock \showarticletitle{Temporal load balancing with service delay
  guarantees for data center energy cost optimization}.
\newblock \bibinfo{journal}{\emph{IEEE Transactions on Parallel and Distributed
  Systems}} \bibinfo{volume}{25}, \bibinfo{number}{3} (\bibinfo{year}{2013}),
  \bibinfo{pages}{775--784}.
\newblock


\bibitem[\protect\citeauthoryear{Maps}{Maps}{2022}]%
        {electricityMap}
\bibfield{author}{\bibinfo{person}{Electricity Maps}.}
  \bibinfo{year}{2022}\natexlab{}.
\newblock \bibinfo{title}{Electricity Maps}.
\newblock
\newblock
\urldef\tempurl%
\url{https://app.electricitymaps.com/map}
\showURL{%
\tempurl}


\bibitem[\protect\citeauthoryear{Masanet, Shehabi, Lei, Smith, and
  Koomey}{Masanet et~al\mbox{.}}{2020}]%
        {Masanet20}
\bibfield{author}{\bibinfo{person}{Eric Masanet}, \bibinfo{person}{Arman
  Shehabi}, \bibinfo{person}{Nuoa Lei}, \bibinfo{person}{Sarah Smith}, {and}
  \bibinfo{person}{Jonathan Koomey}.} \bibinfo{year}{2020}\natexlab{}.
\newblock \showarticletitle{Recalibrating global data center energy-use
  estimates}.
\newblock \bibinfo{journal}{\emph{Science}} \bibinfo{volume}{367},
  \bibinfo{number}{6481} (\bibinfo{year}{2020}), \bibinfo{pages}{984--986}.
\newblock
\showISSN{0036-8075}
\urldef\tempurl%
\url{https://doi.org/10.1126/science.aba3758}
\showDOI{\tempurl}
\showeprint{https://science.sciencemag.org/content/367/6481/984.full.pdf}


\bibitem[\protect\citeauthoryear{Masson-Delmotte, Zhai, P{\"o}rtner, Roberts,
  Skea, Shukla, Pirani, Moufouma-Okia, P{\'e}an, Pidcock,
  et~al\mbox{.}}{Masson-Delmotte et~al\mbox{.}}{2018}]%
        {masson2018global}
\bibfield{author}{\bibinfo{person}{Val{\'e}rie Masson-Delmotte},
  \bibinfo{person}{Panmao Zhai}, \bibinfo{person}{Hans-Otto P{\"o}rtner},
  \bibinfo{person}{Debra Roberts}, \bibinfo{person}{Jim Skea},
  \bibinfo{person}{Priyadarshi~R Shukla}, \bibinfo{person}{Anna Pirani},
  \bibinfo{person}{W Moufouma-Okia}, \bibinfo{person}{C P{\'e}an},
  \bibinfo{person}{R Pidcock}, {et~al\mbox{.}}}
  \bibinfo{year}{2018}\natexlab{}.
\newblock \showarticletitle{Global warming of 1.5 C}.
\newblock \bibinfo{journal}{\emph{An IPCC Special Report on the impacts of
  global warming of}} \bibinfo{volume}{1}, \bibinfo{number}{5}
  (\bibinfo{year}{2018}).
\newblock


\bibitem[\protect\citeauthoryear{McAnany}{McAnany}{2021}]%
        {pjmDR2021}
\bibfield{author}{\bibinfo{person}{James McAnany}.}
  \bibinfo{year}{2021}\natexlab{}.
\newblock \bibinfo{title}{2021 Demand Response Operations Markets Activity
  Report: January 2022}.
\newblock
\newblock
\urldef\tempurl%
\url{https://www.pjm.com/-/media/markets-ops/dsr/2021-demand-response-activity-report.ashx}
\showURL{%
\tempurl}


\bibitem[\protect\citeauthoryear{Menati, Lee, and Xie}{Menati
  et~al\mbox{.}}{2023}]%
        {menati2023modeling}
\bibfield{author}{\bibinfo{person}{Ali Menati}, \bibinfo{person}{Kiyeob Lee},
  {and} \bibinfo{person}{Le Xie}.} \bibinfo{year}{2023}\natexlab{}.
\newblock \showarticletitle{Modeling and analysis of utilizing cryptocurrency
  mining for demand flexibility in electric energy systems: A synthetic texas
  grid case study}.
\newblock \bibinfo{journal}{\emph{IEEE Transactions on Energy Markets, Policy
  and Regulation}} (\bibinfo{year}{2023}).
\newblock


\bibitem[\protect\citeauthoryear{{Microsoft Azure}}{{Microsoft
  Azure}}{[n.\,d.]}]%
        {MicrosoftCloudDatacenters}
\bibfield{author}{\bibinfo{person}{{Microsoft Azure}}.}
  \bibinfo{year}{[n.\,d.]}\natexlab{}.
\newblock \bibinfo{title}{Azure Global Datacenters}.
\newblock
\newblock
\newblock
\shownote{\url{https://azure.microsoft.com/en-us/global-infrastructure/}}.


\bibitem[\protect\citeauthoryear{Miller}{Miller}{2022}]%
        {dominionMoreGreen}
\bibfield{author}{\bibinfo{person}{Rich Miller}.}
  \bibinfo{year}{2022}\natexlab{}.
\newblock \bibinfo{title}{Dominion Energy Plans More Green Power for
  Virginia’s Data Centers}.
\newblock
\newblock
\urldef\tempurl%
\url{https://datacenterfrontier.com/dominion-energy-plans-more-green-power-for-virginias-data-centers/}
\showURL{%
\tempurl}


\bibitem[\protect\citeauthoryear{of~Texas}{of~Texas}{[n.\,d.]}]%
        {LFLTF}
\bibfield{author}{\bibinfo{person}{Electric Reliability~Council of Texas}.}
  \bibinfo{year}{[n.\,d.]}\natexlab{}.
\newblock \bibinfo{title}{Large Flexible Load Task Force}.
\newblock
\newblock
\newblock
\shownote{\url{https://www.ercot.com/committees/tac/lfltf}}.


\bibitem[\protect\citeauthoryear{Papavasiliou and Oren}{Papavasiliou and
  Oren}{2013}]%
        {papavasiliou2013multiarea}
\bibfield{author}{\bibinfo{person}{Anthony Papavasiliou} {and}
  \bibinfo{person}{Shmuel~S Oren}.} \bibinfo{year}{2013}\natexlab{}.
\newblock \showarticletitle{Multiarea stochastic unit commitment for high wind
  penetration in a transmission constrained network}.
\newblock \bibinfo{journal}{\emph{Operations Research}} \bibinfo{volume}{61},
  \bibinfo{number}{3} (\bibinfo{year}{2013}), \bibinfo{pages}{578--592}.
\newblock


\bibitem[\protect\citeauthoryear{Peccarelli}{Peccarelli}{2020}]%
        {Covid-Transform20}
\bibfield{author}{\bibinfo{person}{Brian Peccarelli}.}
  \bibinfo{year}{2020}\natexlab{}.
\newblock \showarticletitle{Three Ways COVID-19 is Accelerating Digital
  Transformation in Professional Services}.
\newblock  (\bibinfo{date}{June} \bibinfo{year}{2020}).
\newblock
\newblock
\shownote{\url{https://bit.ly/34Qitb5}, 37\% growth}.


\bibitem[\protect\citeauthoryear{Pichai}{Pichai}{2019}]%
        {googleRenewPurchase}
\bibfield{author}{\bibinfo{person}{Sundar Pichai}.}
  \bibinfo{year}{2019}\natexlab{}.
\newblock \bibinfo{title}{Our biggest renewable energy purchase ever}.
\newblock
\newblock
\urldef\tempurl%
\url{https://blog.google/outreach-initiatives/sustainability/our-biggest-renewable-energy-purchase-ever/}
\showURL{%
\tempurl}


\bibitem[\protect\citeauthoryear{\protect{New York State Energy Planning
  Board}}{\protect{New York State Energy Planning Board}}{2015}]%
        {NY70by30}
\bibfield{author}{\bibinfo{person}{\protect{New York State Energy Planning
  Board}}.} \bibinfo{year}{2015}\natexlab{}.
\newblock \bibinfo{title}{The Energy to Lead: 2015 New York State Energy Plan}.
\newblock
\newblock
\newblock
\shownote{\url{http://energyplan.ny.gov/Plans/2015.aspx}}.


\bibitem[\protect\citeauthoryear{Pryor, Barthelmie, and Shepherd}{Pryor
  et~al\mbox{.}}{2020}]%
        {pryor202020}
\bibfield{author}{\bibinfo{person}{SC Pryor}, \bibinfo{person}{RJ Barthelmie},
  {and} \bibinfo{person}{TJ Shepherd}.} \bibinfo{year}{2020}\natexlab{}.
\newblock \showarticletitle{20\% of US electricity from wind will have limited
  impacts on system efficiency and regional climate}.
\newblock \bibinfo{journal}{\emph{Scientific reports}} \bibinfo{volume}{10},
  \bibinfo{number}{1} (\bibinfo{year}{2020}), \bibinfo{pages}{1--14}.
\newblock


\bibitem[\protect\citeauthoryear{Qureshi, Weber, Balakrishnan, Guttag, and
  Maggs}{Qureshi et~al\mbox{.}}{2009}]%
        {Maggs09}
\bibfield{author}{\bibinfo{person}{Asfandyar Qureshi}, \bibinfo{person}{Rick
  Weber}, \bibinfo{person}{Hari Balakrishnan}, \bibinfo{person}{John Guttag},
  {and} \bibinfo{person}{Bruce Maggs}.} \bibinfo{year}{2009}\natexlab{}.
\newblock \showarticletitle{Cutting the Electric Bill for Internet-Scale
  Systems}. In \bibinfo{booktitle}{\emph{Proceedings of the ACM SIGCOMM 2009
  Conference on Data Communication}} (Barcelona, Spain)
  \emph{(\bibinfo{series}{SIGCOMM '09})}. \bibinfo{publisher}{Association for
  Computing Machinery}, \bibinfo{address}{New York, NY, USA},
  \bibinfo{pages}{123–134}.
\newblock
\showISBNx{9781605585949}
\urldef\tempurl%
\url{https://doi.org/10.1145/1592568.1592584}
\showDOI{\tempurl}


\bibitem[\protect\citeauthoryear{Radovanovic, Koningstein, Schneider, Chen,
  Duarte, Roy, Xiao, Haridasan, Hung, Care, et~al\mbox{.}}{Radovanovic
  et~al\mbox{.}}{2021}]%
        {radovanovic2021carbon}
\bibfield{author}{\bibinfo{person}{Ana Radovanovic}, \bibinfo{person}{Ross
  Koningstein}, \bibinfo{person}{Ian Schneider}, \bibinfo{person}{Bokan Chen},
  \bibinfo{person}{Alexandre Duarte}, \bibinfo{person}{Binz Roy},
  \bibinfo{person}{Diyue Xiao}, \bibinfo{person}{Maya Haridasan},
  \bibinfo{person}{Patrick Hung}, \bibinfo{person}{Nick Care}, {et~al\mbox{.}}}
  \bibinfo{year}{2021}\natexlab{}.
\newblock \showarticletitle{Carbon-Aware Computing for Datacenters}.
\newblock \bibinfo{journal}{\emph{arXiv preprint arXiv:2106.11750}}
  (\bibinfo{year}{2021}).
\newblock


\bibitem[\protect\citeauthoryear{Roach}{Roach}{2021}]%
        {Microsoft-50DC}
\bibfield{author}{\bibinfo{person}{John Roach}.}
  \bibinfo{year}{2021}\natexlab{}.
\newblock \bibinfo{title}{Microsoft’s virtual datacenter grounds ‘the
  cloud’ in reality}.
\newblock
\newblock
\newblock
\shownote{Microsoft to build 50 to 100 datacenters per year,
  \url{https://news.microsoft.com/innovation-stories/microsofts-virtual-datacenter-grounds-the-cloud-in-reality/}}.


\bibitem[\protect\citeauthoryear{Schwartz, Dodge, Smith, and Etzioni}{Schwartz
  et~al\mbox{.}}{2020}]%
        {GreenAI-CACM20}
\bibfield{author}{\bibinfo{person}{Roy Schwartz}, \bibinfo{person}{Jesse
  Dodge}, \bibinfo{person}{Noah~A. Smith}, {and} \bibinfo{person}{Oren
  Etzioni}.} \bibinfo{year}{2020}\natexlab{}.
\newblock \showarticletitle{Green AI}.
\newblock \bibinfo{journal}{\emph{Commun. ACM}} \bibinfo{volume}{63},
  \bibinfo{number}{12} (\bibinfo{date}{nov} \bibinfo{year}{2020}),
  \bibinfo{pages}{54–63}.
\newblock
\showISSN{0001-0782}
\urldef\tempurl%
\url{https://doi.org/10.1145/3381831}
\showDOI{\tempurl}


\bibitem[\protect\citeauthoryear{Shi, Xu, Zhang, and Wang}{Shi
  et~al\mbox{.}}{2016}]%
        {shi2016leveraging}
\bibfield{author}{\bibinfo{person}{Yuanyuan Shi}, \bibinfo{person}{Bolun Xu},
  \bibinfo{person}{Baosen Zhang}, {and} \bibinfo{person}{Di Wang}.}
  \bibinfo{year}{2016}\natexlab{}.
\newblock \showarticletitle{Leveraging energy storage to optimize data center
  electricity cost in emerging power markets}. In
  \bibinfo{booktitle}{\emph{Proceedings of the Seventh International Conference
  on Future Energy Systems}}. \bibinfo{pages}{1--13}.
\newblock


\bibitem[\protect\citeauthoryear{Staff}{Staff}{2020}]%
        {Covid-Transform20a}
\bibfield{author}{\bibinfo{person}{Staff}.} \bibinfo{year}{2020}\natexlab{}.
\newblock \showarticletitle{COVID-19 Accelerates Cloud Adoption, Market to
  Reach \$1 trillion, IDC}.
\newblock \bibinfo{journal}{\emph{Equipment FA News}} (\bibinfo{date}{October}
  \bibinfo{year}{2020}).
\newblock
\newblock
\shownote{\url{https://www.equipmentfa.com/news/31459/covid-19-accelerates-cloud-adoption-market-to-reach-1t-idc}}.


\bibitem[\protect\citeauthoryear{Stewart, Koenig, Liu, Clausen, Klingert, and
  Bates}{Stewart et~al\mbox{.}}{2019}]%
        {Supercomputer-Unstable-Grid19}
\bibfield{author}{\bibinfo{person}{Grant~L. Stewart},
  \bibinfo{person}{Gregory~A. Koenig}, \bibinfo{person}{Jingjing Liu},
  \bibinfo{person}{Anders Clausen}, \bibinfo{person}{Sonja Klingert}, {and}
  \bibinfo{person}{Natalie Bates}.} \bibinfo{year}{2019}\natexlab{}.
\newblock \showarticletitle{Grid Accommodation of Dynamic HPC Demand}. In
  \bibinfo{booktitle}{\emph{Proceedings of the 48th International Conference on
  Parallel Processing: Workshops}} (Kyoto, Japan) \emph{(\bibinfo{series}{ICPP
  2019})}. \bibinfo{publisher}{Association for Computing Machinery},
  \bibinfo{address}{New York, NY, USA}, Article \bibinfo{articleno}{9},
  \bibinfo{numpages}{4}~pages.
\newblock
\showISBNx{9781450371964}
\urldef\tempurl%
\url{https://doi.org/10.1145/3339186.3339214}
\showDOI{\tempurl}


\bibitem[\protect\citeauthoryear{Sun, Ren, Wu, and Li}{Sun
  et~al\mbox{.}}{2016}]%
        {sun2016online}
\bibfield{author}{\bibinfo{person}{Qihang Sun}, \bibinfo{person}{Shaolei Ren},
  \bibinfo{person}{Chuan Wu}, {and} \bibinfo{person}{Zongpeng Li}.}
  \bibinfo{year}{2016}\natexlab{}.
\newblock \showarticletitle{An online incentive mechanism for emergency demand
  response in geo-distributed colocation data centers}. In
  \bibinfo{booktitle}{\emph{Proceedings of the seventh international conference
  on future energy systems}}. \bibinfo{pages}{1--13}.
\newblock


\bibitem[\protect\citeauthoryear{Tirmazi, Barker, Deng, Haque, Qin, Hand,
  Harchol-Balter, and Wilkes}{Tirmazi et~al\mbox{.}}{2020}]%
        {BorgTNG20}
\bibfield{author}{\bibinfo{person}{Muhammad Tirmazi}, \bibinfo{person}{Adam
  Barker}, \bibinfo{person}{Nan Deng}, \bibinfo{person}{Md~E. Haque},
  \bibinfo{person}{Zhijing~Gene Qin}, \bibinfo{person}{Steven Hand},
  \bibinfo{person}{Mor Harchol-Balter}, {and} \bibinfo{person}{John Wilkes}.}
  \bibinfo{year}{2020}\natexlab{}.
\newblock \showarticletitle{Borg: The next Generation}. In
  \bibinfo{booktitle}{\emph{Proceedings of the Fifteenth European Conference on
  Computer Systems}} (Heraklion, Greece) \emph{(\bibinfo{series}{EuroSys
  '20})}. \bibinfo{publisher}{Association for Computing Machinery},
  \bibinfo{address}{New York, NY, USA}, Article \bibinfo{articleno}{30},
  \bibinfo{numpages}{14}~pages.
\newblock
\showISBNx{9781450368827}
\urldef\tempurl%
\url{https://doi.org/10.1145/3342195.3387517}
\showDOI{\tempurl}


\bibitem[\protect\citeauthoryear{{United Nations Framework Convention on
  Climate Change}}{{United Nations Framework Convention on Climate
  Change}}{2015}]%
        {Paris2015}
\bibfield{author}{\bibinfo{person}{{United Nations Framework Convention on
  Climate Change}}.} \bibinfo{year}{2015}\natexlab{}.
\newblock \bibinfo{title}{Paris Climate Change Conference}.
\newblock
\newblock
\newblock
\shownote{\url{http://unfccc.int/meetings/paris_nov_2015/meeting/8926.php}}.


\bibitem[\protect\citeauthoryear{Urgaonkar, Urgaonkar, Neely, and
  Sivasubramaniam}{Urgaonkar et~al\mbox{.}}{2011}]%
        {urgaonkar2011optimal}
\bibfield{author}{\bibinfo{person}{Rahul Urgaonkar}, \bibinfo{person}{Bhuvan
  Urgaonkar}, \bibinfo{person}{Michael~J Neely}, {and} \bibinfo{person}{Anand
  Sivasubramaniam}.} \bibinfo{year}{2011}\natexlab{}.
\newblock \showarticletitle{Optimal power cost management using stored energy
  in data centers}. In \bibinfo{booktitle}{\emph{Proceedings of the ACM
  SIGMETRICS joint international conference on Measurement and modeling of
  computer systems}}. \bibinfo{pages}{221--232}.
\newblock


\bibitem[\protect\citeauthoryear{Vanian and Leswing}{Vanian and
  Leswing}{2023}]%
        {genAICost}
\bibfield{author}{\bibinfo{person}{Jonathan Vanian} {and} \bibinfo{person}{Kif
  Leswing}.} \bibinfo{year}{2023}\natexlab{}.
\newblock \bibinfo{title}{ChatGPT and generative AI are booming, but the costs
  can be extraordinary}.
\newblock
\newblock
\urldef\tempurl%
\url{https://www.cnbc.com/2023/03/13/chatgpt-and-generative-ai-are-booming-but-at-a-very-expensive-price.html}
\showURL{%
\tempurl}


\bibitem[\protect\citeauthoryear{Wacket}{Wacket}{2021}]%
        {germanyPhaseOutCoal}
\bibfield{author}{\bibinfo{person}{Markus Wacket}.}
  \bibinfo{year}{2021}\natexlab{}.
\newblock \bibinfo{title}{German parties agree on 2030 coal phase-out in
  coalition talks}.
\newblock
\newblock
\urldef\tempurl%
\url{https://www.reuters.com/business/cop/exclusive-germanys-government-in-waiting-agrees-phase-out-coal-by-2030-sources-2021-11-23/}
\showURL{%
\tempurl}


\bibitem[\protect\citeauthoryear{WattTime}{WattTime}{2022}]%
        {wattTimeAER}
\bibfield{author}{\bibinfo{person}{WattTime}.} \bibinfo{year}{2022}\natexlab{}.
\newblock \bibinfo{title}{Automated Emissions Reduction}.
\newblock
\newblock
\urldef\tempurl%
\url{https://www.watttime.org/solutions/automated-emissions-reduction-aer/}
\showURL{%
\tempurl}


\bibitem[\protect\citeauthoryear{Yang and Chien}{Yang and Chien}{2016}]%
        {YangChien-IPDPS16}
\bibfield{author}{\bibinfo{person}{Fan Yang} {and} \bibinfo{person}{Andrew~A
  Chien}.} \bibinfo{year}{2016}\natexlab{}.
\newblock \showarticletitle{ZCCloud: Exploring wasted green power for
  high-performance computing}. In \bibinfo{booktitle}{\emph{2016 IEEE
  International Parallel and Distributed Processing Symposium (IPDPS)}}. IEEE,
  \bibinfo{pages}{1051--1060}.
\newblock


\bibitem[\protect\citeauthoryear{Yang and Chien}{Yang and Chien}{2017}]%
        {YangChien-TPDS17}
\bibfield{author}{\bibinfo{person}{Fan Yang} {and} \bibinfo{person}{Andrew~A.
  Chien}.} \bibinfo{year}{2017}\natexlab{}.
\newblock \showarticletitle{Large-scale and Extreme-Scale Computing with
  Stranded Green Power: Opportunities and Costs}.
\newblock \bibinfo{journal}{\emph{IEEE Transactions on Parallel and Distributed
  Systems}} \bibinfo{volume}{29}, \bibinfo{number}{5} (\bibinfo{date}{December}
  \bibinfo{year}{2017}).
\newblock


\bibitem[\protect\citeauthoryear{Yang, Hajiesmaili, Sitaraman, Mallada, Wong,
  and Wierman}{Yang et~al\mbox{.}}{2019}]%
        {Lin19}
\bibfield{author}{\bibinfo{person}{Lin Yang}, \bibinfo{person}{Mohammad~Hassan
  Hajiesmaili}, \bibinfo{person}{Ramesh~K. Sitaraman}, \bibinfo{person}{Enrique
  Mallada}, \bibinfo{person}{Wing~Shing Wong}, {and} \bibinfo{person}{Adam
  Wierman}.} \bibinfo{year}{2019}\natexlab{}.
\newblock \showarticletitle{Online Inventory Management with Application to
  Energy Procurement in Data Centers}.
\newblock \bibinfo{journal}{\emph{CoRR}}  \bibinfo{volume}{abs/1901.04372}
  (\bibinfo{year}{2019}).
\newblock
\showeprint[arxiv]{1901.04372}
\urldef\tempurl%
\url{http://arxiv.org/abs/1901.04372}
\showURL{%
\tempurl}


\bibitem[\protect\citeauthoryear{Zhang}{Zhang}{2022}]%
        {Zhang2022thesis}
\bibfield{author}{\bibinfo{person}{Chaojie Zhang}.}
  \bibinfo{year}{2022}\natexlab{}.
\newblock \emph{\bibinfo{title}{Eliminating the Capacity Variation Penalty for
  Cloud Resource Management}}.
\newblock \bibinfo{thesistype}{Ph.\,D. Dissertation}. \bibinfo{school}{The
  University of Chicago}, \bibinfo{address}{Chicago, IL, USA}.
\newblock Advisor(s) Chien, Andrew A.
\newblock


\bibitem[\protect\citeauthoryear{Zhang and Chien}{Zhang and Chien}{2021}]%
        {zhang2021scheduling}
\bibfield{author}{\bibinfo{person}{Chaojie Zhang} {and}
  \bibinfo{person}{Andrew~A Chien}.} \bibinfo{year}{2021}\natexlab{}.
\newblock \showarticletitle{Scheduling Challenges for Variable Capacity
  Resources}. In \bibinfo{booktitle}{\emph{Workshop on Job Scheduling for
  Parallel Processing (JSSPP)}}.
\newblock


\bibitem[\protect\citeauthoryear{Zhang, Ren, Wu, and Li}{Zhang
  et~al\mbox{.}}{2015}]%
        {zhang2015truthful}
\bibfield{author}{\bibinfo{person}{Linquan Zhang}, \bibinfo{person}{Shaolei
  Ren}, \bibinfo{person}{Chuan Wu}, {and} \bibinfo{person}{Zongpeng Li}.}
  \bibinfo{year}{2015}\natexlab{}.
\newblock \showarticletitle{A truthful incentive mechanism for emergency demand
  response in colocation data centers}. In \bibinfo{booktitle}{\emph{2015 IEEE
  Conference on Computer Communications (INFOCOM)}}. IEEE,
  \bibinfo{pages}{2632--2640}.
\newblock


\bibitem[\protect\citeauthoryear{Zhang, Roald, Chien, Birge, and Zavala}{Zhang
  et~al\mbox{.}}{2020}]%
        {zhang2020flexibility}
\bibfield{author}{\bibinfo{person}{Weiqi Zhang}, \bibinfo{person}{Line~A
  Roald}, \bibinfo{person}{Andrew~A Chien}, \bibinfo{person}{John~R Birge},
  {and} \bibinfo{person}{Victor~M Zavala}.} \bibinfo{year}{2020}\natexlab{}.
\newblock \showarticletitle{Flexibility from networks of data centers: A market
  clearing formulation with virtual links}.
\newblock \bibinfo{journal}{\emph{Electric Power Systems Research}}
  \bibinfo{volume}{189} (\bibinfo{year}{2020}), \bibinfo{pages}{106723}.
\newblock


\bibitem[\protect\citeauthoryear{Zhang, Wang, and Wang}{Zhang
  et~al\mbox{.}}{2012}]%
        {zhang2012electricity}
\bibfield{author}{\bibinfo{person}{Yanwei Zhang}, \bibinfo{person}{Yefu Wang},
  {and} \bibinfo{person}{Xiaorui Wang}.} \bibinfo{year}{2012}\natexlab{}.
\newblock \showarticletitle{Electricity bill capping for cloud-scale data
  centers that impact the power markets}. In \bibinfo{booktitle}{\emph{2012
  41st International Conference on Parallel Processing}}. IEEE,
  \bibinfo{pages}{440--449}.
\newblock


\bibitem[\protect\citeauthoryear{Zheng, Chien, and Suh}{Zheng
  et~al\mbox{.}}{2020}]%
        {Chien-Joule20}
\bibfield{author}{\bibinfo{person}{Jiajia Zheng}, \bibinfo{person}{Andrew~A.
  Chien}, {and} \bibinfo{person}{Sangwon Suh}.}
  \bibinfo{year}{2020}\natexlab{}.
\newblock \showarticletitle{Mitigating Curtailment and Carbon Emissions through
  Load Migration between Data Centers}.
\newblock \bibinfo{journal}{\emph{Joule}} (\bibinfo{date}{October}
  \bibinfo{year}{2020}).
\newblock
\showISSN{2542-4351}
\urldef\tempurl%
\url{https://doi.org/10.1016/j.joule.2020.08.001}
\showDOI{\tempurl}


\bibitem[\protect\citeauthoryear{Zhou, Liu, Chen, and Li}{Zhou
  et~al\mbox{.}}{2018}]%
        {zhou2018truthful}
\bibfield{author}{\bibinfo{person}{Zhi Zhou}, \bibinfo{person}{Fangming Liu},
  \bibinfo{person}{Shutong Chen}, {and} \bibinfo{person}{Zongpeng Li}.}
  \bibinfo{year}{2018}\natexlab{}.
\newblock \showarticletitle{A truthful and efficient incentive mechanism for
  demand response in green datacenters}.
\newblock \bibinfo{journal}{\emph{IEEE Transactions on Parallel and Distributed
  Systems}} \bibinfo{volume}{31}, \bibinfo{number}{1} (\bibinfo{year}{2018}),
  \bibinfo{pages}{1--15}.
\newblock


\bibitem[\protect\citeauthoryear{Zhou, Liu, Li, Li, Jin, Zou, and Liu}{Zhou
  et~al\mbox{.}}{2014}]%
        {zhou2014fuel}
\bibfield{author}{\bibinfo{person}{Zhi Zhou}, \bibinfo{person}{Fangming Liu},
  \bibinfo{person}{Bo Li}, \bibinfo{person}{Baochun Li}, \bibinfo{person}{Hai
  Jin}, \bibinfo{person}{Ruolan Zou}, {and} \bibinfo{person}{Zhiyong Liu}.}
  \bibinfo{year}{2014}\natexlab{}.
\newblock \showarticletitle{Fuel cell generation in geo-distributed cloud
  services: A quantitative study}. In \bibinfo{booktitle}{\emph{2014 IEEE 34th
  International Conference on Distributed Computing Systems}}. IEEE,
  \bibinfo{pages}{52--61}.
\newblock


\bibitem[\protect\citeauthoryear{Zhou, Liu, Zou, Liu, Xu, and Jin}{Zhou
  et~al\mbox{.}}{2016}]%
        {zhou2016carbon}
\bibfield{author}{\bibinfo{person}{Zhi Zhou}, \bibinfo{person}{Fangming Liu},
  \bibinfo{person}{Ruolan Zou}, \bibinfo{person}{Jiangchuan Liu},
  \bibinfo{person}{Hong Xu}, {and} \bibinfo{person}{Hai Jin}.}
  \bibinfo{year}{2016}\natexlab{}.
\newblock \showarticletitle{Carbon-Aware Online Control of Geo-Distributed
  Cloud Services}.
\newblock \bibinfo{journal}{\emph{IEEE Transactions on Parallel and Distributed
  Systems}} \bibinfo{volume}{27}, \bibinfo{number}{9} (\bibinfo{year}{2016}),
  \bibinfo{pages}{2506--2519}.
\newblock


\end{thebibliography}

\appendix
%\textbf{These are fragmented and appear to be in the wrong order? and what about Section 5.3?}

\section{Direct-current Optimal Power Flow Formulation}
We model the grid operation using the direct-current optimal power flow (DC-OPF) model. Notations in this model are listed below:

\label{appendix:dc-opf}
\begin{table}[H]
\caption{DC-OPF Notation: Sets}
\label{tab:notation-set}
\begin{center}
\scalebox{0.85}{
\begin{tabular}{p{35pt}|p{85pt}|p{35pt}|p{85pt}}
  \hline
  Notation & Description & Notation & Description\\
  \hline
  $\mathcal{D}\ (\mathcal{D}_n)$ & Demand loads (at bus $n$) & $\mathcal{G}\ (\mathcal{G}_n)$ & Generators (at bus $n$)\\
  $\mathcal{I}\ (\mathcal{I}_n)$ & Import points (at bus $n$) & $\mathcal{L}$ & Transmission lines\\
  $\mathcal{L}_n^+/\mathcal{L}_n^-$ & Transmission lines to/from bus $n$ & $\mathcal{N}$ & Buses\\
  $\mathcal{R}\ (\mathcal{R}_n)$ & Renewable generators (at bus $n$) & $\mathcal{T}$ & Time periods\\
  $\mathcal{W}\ (\mathcal{W}_n)$ & Wind farms (at bus $n$) & $DC (DC_n)$ & Datacenters (at bus $n$)\\
  \hline
\end{tabular}
}
\end{center}
\end{table}

\begin{table}[H]
\caption{DC-OPF Notation: Parameters}
\label{tab:notation-param}
\begin{center}
\scalebox{0.85}{
\begin{tabular}{p{35pt}|p{85pt}|p{35pt}|p{85pt}}
  \hline
  Notation & Description & Notation & Description\\
  \hline
  $B_l$ & Susceptance of transmission line $l$ & $C_i$ & Generation cost of generator $i$\\
  $C_j^d$ & Load-shedding penalty at load $j$ & $C_i^w$ & Curtailment penalty at wind farm $i$\\
  $C_i^m$ & Curtailment penalty at import point $i$ & $C_i^r$ & Curtailment penalty at renewable $i$\\
  $D_{j,t}$ & Demand load of consumer $j$ at time $t$ & $F^{max}_l$ & Maximum power flow of transmission line $l$\\
  $M_{i,t}$ & Power production of import $i$ at time $t$ & $P^{max}_i$ & Maximum power output of generator $i$\\
  $R_{i,t}$ & Power production of renewable $i$ at time $t$ & $RU_i$ & Ramp-up limit of generator $i$\\
  $RD_i$ & Ramp-down limit of generator $i$ & $W_{w,t}$ & Power from wind farm $w$ at time $t$\\
  $\Theta_{n,t}^{min}$ & Minimum phase angle at bus $n$ at time $t$ & $\Theta_{n,t}^{max}$ & Maximum phase angle at bus $n$ at time $t$\\
  \hline
\end{tabular}
}
\end{center}
\end{table}

\begin{table}[H]
\caption{DC-OPF Notation: Decision Variables}
\label{tab:notation-dv}
\begin{center}
\scalebox{0.85}{
\begin{tabular}{p{35pt}|p{85pt}|p{35pt}|p{85pt}}
  \hline
  Notation & Description & Notation & Description\\
  \hline
  $d_{j,t}$ & Load shedding at load $j$ at time $t$ & $f_{l,t}$ & Power flow of line $l$ at time $t$\\
  $m_{i,t}$ & Curtailment at import $i$ at time $t$ & $p_{i,t}$ & Power from generator $i$ at time $t$\\
  $r_{i,t}$ & Curtailment at renewable $i$ at time $t$ & $w_{i,t}$ & Curtailment at wind farm $i$ at time $t$\\
  $\theta_{n,t}$ & Phase angle at bus $n$ at time $t$ &  & \\
  \hline
\end{tabular}
}
\end{center}
\end{table}

Datacenter capacity levels $cap_{i,t}$ are external decisions by the datacenters or coordinator(s). The power grid solves the DC-OPF model (one-day time horizon with hourly intervals in our studies), minimizing the dispatch cost (\ref{eq:ED:obj}) that consists of generation cost, load shedding penalty, and curtailment penalties:
\begin{subequations}
\label{eq:ED}
\begin{align}
  \min \quad 
  & \sum_{t\in\mathcal{T}} \left( \sum_{i\in\mathcal{G}} C_i p_{i,t} + \sum_{j\in\mathcal{D}} C_j^d d_{j,t} + \sum_{i\in\mathcal{I}} C_i^m m_{i,t} \right. \notag \\
  & \qquad \left. + \sum_{i\in\mathcal{W}} C_i^w w_{i,t} + \sum_{i\in\mathcal{R}} C_i^r r_{i,t} \right) \label{eq:ED:obj} \\
  \text{s.t.} \quad
  & \sum_{l\in\mathcal{L}_n^+} f_{l,t} - \sum_{l\in\mathcal{L}_n^-} f_{l,t} + \sum_{i\in\mathcal{G}_n} p_{i,t} + \sum_{i\in\mathcal{I}_n} (M_{i,t} - m_{i,t}) \notag \\
  & + \sum_{i\in\mathcal{W}_n} (W_{i,t} - w_{i,t}) + \sum_{i\in\mathcal{R}_n} (R_{i,t} - r_{i,t}) \notag \\
  & = \sum_{j\in\mathcal{D}_n} (D_{j,t} - d_{j,t}) + \sum_{i\in DC_n}cap_{i,t}, \quad \forall n\in\mathcal{N}, t\in\mathcal{T}, \label{eq:ED:flow} \\
  & f_{l,t} = B_l (\theta_{n,t} - \theta_{m,t}), \quad \forall l=(m,n)\in\mathcal{L}, t\in\mathcal{T}, \label{eq:ED:angle}\\ 
  & -F_l^{max} \leq f_{l,t} \leq F_l^{max}, \quad \forall l\in\mathcal{L}, t\in\mathcal{T}, \label{eq:ED:linecap}\\
  & \Theta_n^{min} \leq \theta_{n,t} \leq \Theta_n^{max} \quad \forall n\in\mathcal{N}, t\in\mathcal{T}, \label{eq:ED:anglerange} \\
  & -RD_i \leq p_{i,t} - p_{i,t-1} \leq RU_i, \quad \forall i\in\mathcal{G}, t\in\mathcal{T}, \label{eq:ED:ramp} \\
  & 0 \leq p_{i,t} \leq P_i^{max}, \quad \forall i\in\mathcal{G}, t\in\mathcal{T}, \label{eq:ED:maxgen}\\
  & 0 \leq d_{j,t} \leq D_{j,t}, \quad \forall j\in\mathcal{D}, t\in\mathcal{T}, \label{eq:ED:maxshed} \\
  & 0 \leq m_{i,t} \leq M_{j,t}, \quad \forall i\in\mathcal{I}, t\in\mathcal{T}, \label{eq:ED:maximport} \\
  & 0 \leq w_{i,t} \leq W_{j,t}, \quad \forall i\in\mathcal{W}, t\in\mathcal{T}, \label{eq:ED:maxwind} \\
  & 0 \leq r_{i,t} \leq R_{j,t}, \quad \forall i\in\mathcal{R}, t\in\mathcal{T}. \label{eq:ED:maxrenewable}
\end{align}
\end{subequations}

The generation sources include conventional thermal power plants (e.g. gas, nuclear, coal), non-wind renewables (e.g. hydro), imports, and wind power plants. Due to long-term commitments for imports or goal of reducing carbon emissions with renewables, the imports and renewables are non-dispatchable as in \cite{KYZC2016} but can be curtailed at the cost of $C_i^m$ and $C_i^r$ \$/MWh ($C_i^w$ \$/MWh for wind) respectively. Sometimes the power supply may not meet the demand, and each unit of load shedding (not served load) is at the cost of value of lost load (VOLL) $C_j^d$. In this paper, the unit generation cost is 1/2/4 \$/MWh for nuclear/coal/gas. The penalties are 500/100/1,000 \$/MWh for import/wind/non-wind renewables curtailment, 1,000 \$/MWh for load shedding.

The constraints are typical for DC-OPF. Constraint \ref{eq:ED:flow} represents the balance constraint at each bus, whose associated dual value is the locational marginal price (LMPrice) at that bus, indicating marginal cost of adding 1 MW load at a specific location, so the price can go negative/high when curtailment/load shedding happens. Constraints \ref{eq:ED:angle}--\ref{eq:ED:anglerange} represent how the power flow (\ref{eq:ED:angle}) is determined given the line capacity (\ref{eq:ED:linecap}) and phase angle (\ref{eq:ED:anglerange}) limits. Constraint \ref{eq:ED:ramp} limits conventional power plants' rate of  ramping up/down generation. Constraints \ref{eq:ED:maxgen}--\ref{eq:ED:maxrenewable} bound the conventional generation, load shedding, and curtailments (\ref{eq:ED:maximport}--\ref{eq:ED:maxrenewable}) respectively.

\section{Fuel Carbon Emission Rates}
\label{sec:fuelCI}
Below are the fuel carbon emission rates (carbon emissions per MWh energy generated from that fuel) we use to calculate carbon emissions: 
\begin{table}[H]
\caption{Carbon Emission Rates of Different Fuels \cite{eGrid, caisoGHG}}
\label{tab:appendix_carbon}
\begin{tabular}{cc}
  \hline
  Generation Type & Carbon Emission Rate (kg CO$_2$/MWh)\\
  \hline
  Coal & 895.2\\
  Natural Gas & 388.9\\
  Oil & 877.6\\
  Dual-fuel & 633.3\\
  Nuclear & 0\\
  Geothermal & 107.6\\
  Biomass & 0\\
  Hydro & 0\\
  Wind & 0\\
  Import & 428\\
  \hline
\end{tabular}
\end{table}

\end{document}